\documentclass[preprint,12pt]{elsarticle}

\usepackage{amssymb}
\usepackage{amsmath}
\usepackage{bm}
\usepackage[colorlinks=true, allcolors=blue]{hyperref}

\newcommand{\pd}[2]{\frac{\partial #1}{\partial #2}}
\newcommand{\sgn}[1]{\text{sgn}\left(#1\right)}
\newcounter{bla}

\journal{Computer Physics Communications}

\begin{document}

\begin{frontmatter}



\title{GaDE - GPU-acceleration of time-dependent Dirac Equation for exascale}


\author[a]{Johanne Elise Vembe} 
\author[b,c]{Marcin Krotkiewski}
\author[d,c]{Magnar Bjørgve}
\author[a]{Morten Førre}
\author[e,c]{Hicham Agueny\textsuperscript{*}}

\cortext[author] {Corresponding author.\\\textit{E-mail address:} hicham.agueny@uib.no}
\address[a]{Department of Physics and Technology, University of Bergen, 5020 Bergen, Norway}
\address[b]{USIT, University of Oslo, 0373 Oslo, Norway}
\address[c]{Norwegian Research Infrastructure Services - NRIS}
\address[d]{HPC group, Enterprise Digital Services for Research and Dissemination. Department of IT, The Arctic University of Norway, 9037 Tromso, Norway}
\address[e]{Information Technology Department, University of Bergen, 5020 Bergen, Norway}

\begin{abstract}
Modern heterogeneous high-performance computing (HPC) systems powered by advanced graphics processing unit (GPU) architectures enable accelerating computing with unprecedented performance and scalability. Here, we present a GPU-accelerated solver for the three-dimensional (3D) time-dependent Dirac Equation optimized for distributed HPC systems. The solver named GaDE is designed to simulate the electron dynamics in atoms induced by electromagnetic fields in the relativistic regime. It combines MPI with CUDA/HIP to target both NVIDIA and AMD GPU architectures. We discuss our implementation strategies in which most of the computations are carried out on GPUs, taking advantage of the GPU-aware MPI feature to optimize communication performance. We evaluate GaDE on the pre-exascale supercomputers, LUMI, powered by AMD MI250X GPU and the HPE's Slingshot interconnect. Single GPU performance on NVIDIA A100, GH200 and AMD MI250X shows comparable performance between A100 and MI250X in compute and memory bandwidth, with GH200 delivering higher performance. Weak scaling on LUMI demonstrates exceptional scalability, achieving 85\% parallel efficiency across 2048 GPUs, while strong scaling delivers a 16× speedup on 32 GPUs - 50\% efficiency for a communication-intensive, time-dependent Dirac equation solver. These results demonstrate GaDE’s high scalability, making it suitable for exascale systems and enabling predictive simulations for ultra-intense laser experiments probing relativistic quantum effects.
\end{abstract}


\end{frontmatter}


\begin{small}
\noindent
{\bf PROGRAM SUMMARY} \\[4pt]
{\em Program Title:} GaDE \\
{\em Developer's repository link:} \url{https://github.com/JVembe/dirac_hydrogen_code} \\
{\em Licensing provisions:} GPLv3 \\
{\em Programming language:} C++, C
\end{small}

\section{INTRODUCTION}
\label{intro}
Advances in ultrafast laser technology have opened
the possibility of visualizing and controlling the matter at the level of electrons \cite{Goulielmakis2010,Sugioka2014}. Such progress has been guided by theoretical developments, which have been extensively focused on a non-relativistic quantum description of the dynamics of electrons induced by laser fields. This has been widely studied by solving the time-dependent Schr\"odinger equation (TDSE) covering a broad wavelength range (see \cite{Johanne2024,Agueny2018,Chovancova2017} and references therein).

However, a comprehensive picture of this dynamics in the relativistic regime is still missing, especially in the long-wavelength region and for large quantum systems. This requires solving the time-dependent Dirac equation (TDDE), which provides an accurate relativistic description of the electron dynamics induced by electromagnetic fields. It is well-recognized that solving the TDDE presents some numerical challenges (see e.g. \cite{Stacey1982,Muller1998, 
Braun1999,FillionGourdeau2012,FillionGourdeau2014}). This is mainly due to numerical artifact caused by (i) fermion doubling problem arising from the discretization of the derivative operator. This makes it difficult to ensure conservation of the current density on the grid \cite{Maquet2002,Muller1998,Stacey1982}. (ii) Simulation in a finite box, which requires careful consideration of the spurious reflection of the time-dependent wavefunction on the edges, for example by use of absorbing potentials \cite{Maquet2002,ANTOINE2014268,Antoine2017_absorb}. (iii) Treatment of potentials with singularities, in particular the Coulomb potential. All of these issues play a role in determining which discretization methods are feasible in solving the TDDE.

On the time evolution side, solving the TDDE additionally requires refined temporal and spatial grids, due to the presence of negative energy solutions whose influence on the dynamics of the electron are known as \textit{zitterbewegung} \cite{Ivanov2015}, a rapid oscillatory motion of the electron occurring on the Compton time scale, $t \propto \frac{\hbar}{m c^2}$. The need to resolve these phenomena renders the numerical solution computationally intensive for laser-matter interaction, as these processes take place at a significantly larger time scale than the zitterbewegung. Addressing these issues requires combining sophisticated numerical schemes with computational models that utilize advanced computer architecture.

So far, existing Dirac-based solvers are mostly limited to a serial implementation of the time-independent Dirac equation developed for atomic structure calculations \cite{Certik2013,Certik2024} and to a parallel implementation of the TDDE that utilizes the Message Passing Interface (MPI) library \cite{Antoine2017,FillionGourdeau2012,Beerwerth2015}. In these cases, the performance analysis was performed solely on strong scaling and that the solver's scalability was limited to a maximum of 256 MPI-processes (e.g. in \cite{Antoine2017} the reported parallel efficiency was 25 \% for the 3D-model). Here, although strong scaling measures the parallel efficiency of a solver for a fixed problem size as the number of processes increases, it does not provide a complete picture of a solver's scalability, especially when both problem size and computational resources grow proportionally. 

On the other hand, there has been an attempt to utilize the graphics processing unit (GPU) to solve the TDDE by implementing the CUDA programming model \cite{Bauke2011}. However, the implementation was limited to a two-dimensional (2D) scheme of the TDDE, and the solver was designed specifically to target a single GPU. Therefore, there is a need for developing optimized TDDE-based solvers to leverage modern heterogeneous high-performance computing (HPC) systems; thus guiding the scientific community towards conducting accurate simulations in the relativistic regime. This in turn will help explore new physics to advance our understanding of the relativistic light-matter interaction.

In recent years, there has been great interest in utilizing GPU accelerators for high-performance computing \cite{Li2025,Yeung2025,Sathyanarayana2025,Budiardja2023,Clay2018}. These graphics accelerators have the potential to accelerate computing significantly over traditional central processing units (CPUs) (see e.g. \cite{Bauke2011,Friedrichs2009}). This is primarily attributed to the massive processing capability of the GPU accelerators designed with a vast array of parallel processing units, which make them ideal for high throughput computing. The latter has been identified as an important factor for future design of hardware architecture \cite{Asanovic2006}. And because of their computing power, GPUs are integrated into HPC systems. Currently, the world's 5 most powerful supercomputers are powered by GPUs \cite{Top500}, indicating their essential role in exploring the performance and scalability of scientific applications to improve their efficiency. For instance, a GPU-enabled simulation of 3D turbulence with a resolution of up to 35 trillion grid points has been reported to be performed on up to 4096 compute nodes of \textit{Frontier}, the world's second exascale supercomputer \cite{Yeung2025}. Other GPU-enabled simulations of turbulent flows have been executed on up to 256 and 512 compute nodes, respectively, on the two largest EuroHPC pre-exascale systems, LUMI and Leonardo \cite{Sathyanarayana2025}. It therefore becomes apparent that developing solvers that effectively harness the power of GPU accelerators in large HPC systems can significantly enhance computational efficiency and performance.

Motivated by the general interest of studying the quantum electron motion in the relativistic regime and by the highly parallel architecture and high-throughput of GPU accelerators, we develop a GPU-based solver named GaDE to simulate the 3D-TDDE. The implementation combines MPI and the CUDA/HIP programming model and takes advantage of the GPU-aware MPI feature to further enhance performance by reducing latency \cite{Shainer2011}. Such a combination offers the potential to utilize multiple GPUs across multiple nodes in modern heterogeneous HPC systems. The solver GaDE is portable and can be compiled on both AMD GPUs and NVIDIA GPUs. Our implementation is designed to utilize efficient numerical schemes that combine the use of B-spline basis functions to describe the atomic structure, and a time-propagation scheme based on Crank-Nicholson. This is in addition to open-source libraries optimized for GPU accelerators.

We carry out experiments on one of the top 10 pre-exascale supercomputers named LUMI at CSC's data center \cite{Lumi,LumiGuide}. The supercomputer LUMI is powered by AMD GPUs. Note that so far, five of the top 10 supercomputers are powered by AMD GPUs \cite{Top500}. Here, we evaluate GPU-related metrics of our solver, mainly, bandwidth, strong scaling and weak scaling on up to 2048 GPUs. We initially run an experiment targeting a single GPU performance covering advanced GPUs architectures, specifically, NVIDIA A100, NVIDIA GH200 and AMD MI250X. We further extend the experiment to evaluate the scalability of the solver on multiple GPUs across multiple nodes on the supercomputer LUMI. Overall, our findings indicate excellent efficiency of 96\% on 2048 GPUs. 

Due to its wavelength-agnostic model of the ionization process, the solver has the ability to solve the 3D TDDE in a wide range of wavelengths of the laser field, although development has targeted performance in the ultrafast regime. Taking advantage of modern HPC systems, this will open opportunities for uncovering quantum phenomena in the relativistic regime that are so far hindered by the lack of optimized Dirac solvers designed for large-scale computations. Our solver therefore will add new insights into attosecond and strong-field physics and towards zeptosecond physics \cite{Grundmann2020,Adnani2022} and coherent control in the relativistic regime. The GaDE solver is available as open-source code from the GitHub repository \cite{GitHub}.

The paper is organized as follows. In Secs. \ref{theory} and \ref{computing}, we provide the theoretical model and
the computational basis to solve the 3D-TDDE. This includes a description of the used algorithms, the GPU implementation and the solver's structure. Section \ref{result} is devoted to the performance analysis, and particularly, the evaluation of the GPU memory bandwidth, and strong scaling and weak scaling. Finally, conclusions are given in Sec. \ref{conclusion}. 

\section{Theory and model}\label{theory}

The theoretical model underlying the GaDE solver has previously been outlined in \cite{Johanne2024}, and its most relevant features are given here for convenience. In the following, we use atomic unit (a.u.), with $e=m_e=4\pi\epsilon_0=\hbar=1$, unless otherwise stated.

\subsection{Dirac equation}
The TDDE governing the relativistic quantum dynamics of spin-1/2 particles in electromagnetic fields can be written as
\begin{eqnarray}\label{eqn:TDDE}
    i \pd{}{t} \psi(t) = H(t) \psi(t).
\end{eqnarray}
The Dirac wavefunction $\psi(t)$ here is a four-component bispinor, representing the state of the system at a given point in time.
$H(t)$ is the Dirac Hamiltonian, which can be separated into stationary $H_0$ and time-dependent $H_I(t)$ components 
\begin{eqnarray}
    H(t) = H_0 + H_I(t)\\
    H_0 = \beta c^2 - I_4 V(r) + c \bm{\alpha} \cdot \bm{p} \\
    \label{eqn:interactionH}
    H_I(t) = - c \bm{\alpha} \cdot \bm{A}(t,\bm{r}),
\end{eqnarray}
where $\bm\alpha$ and $\beta$ are the Dirac matrices
\begin{eqnarray}
    \beta = \begin{pmatrix}
        0 && 0\\
        0 && -2I_2
    \end{pmatrix}, 
    \alpha_i = \begin{pmatrix}
        0 && \sigma_i \\
        -\sigma_i && 0
    \end{pmatrix}.
\end{eqnarray}
Here, the choice of a nonstandard $\beta$ matrix is made to shift the energy levels downwards by $c^2$ for ease of comparison with nonrelativistic models.
$\bm{p}$ is the vector of momentum operators:
\begin{eqnarray}
    \bm{p} = i \bm{\nabla}
\end{eqnarray}
$\bm{A}$ is the external electromagnetic potential representing the incident laser pulse, $V(r)$ is the Coulomb potential, and $I_n$ is the $n\times n$ identity matrix. $c$ is the speed of light in vacuum.

\subsection{Numerical scheme}

As the code is designed primarily for atomic systems, we numerically solve the TDDE in Eq.(\ref{eqn:TDDE}) in spherical coordinates by expanding the time-dependent wave function $\psi$ \cite{Johanne2024},
\begin{eqnarray}\label{eqn:basis}
    \psi(\bm r, t) = \sum_{n,\kappa,\mu} c_{n,\kappa,\mu}(t) r^{-1}
    \begin{pmatrix}
        P_{k,\kappa}(r)X_{\kappa,\mu}(\bm{\hat{r}}) \\ 
        iQ_{k,\kappa}(r)X_{-\kappa,\mu}(\bm{\hat{r}})
    \end{pmatrix}
\end{eqnarray}
Here, the functions $P$ and $Q$ are the solution of the radial part of the Dirac equation, while $X_{\kappa,\mu}$ are the spherical bispinors representing angular momentum states in a relativistic framework. $c_{n,\kappa,\mu}$ is a set of expansion coefficients representing the wavefunction in this basis, which is stored in memory during the code execution. The unit vector $\bm{\hat{r}}$ represents the direction of $\bm{r}$.

The possibility of using angular momentum eigenstates for the basis is an important reason for using spherical coordinates, as they allow us to make use of sparse, highly restrictive selection rules for ionization processes. This choice carries with it some difficulties in handling the linearly polarized laser pulse propagating along a Cartesian direction, as will be shown later in this chapter.

The spherical bispinors are expressed in terms of spherical harmonics
\begin{eqnarray}\label{eqn:spinorbasis}
    X_{\kappa,\mu}(\bm{\hat{r}}) = 
    \begin{pmatrix}
        C_{\mu-\frac{1}{2}, \frac{1}{2},\mu}^{l_\kappa,\frac{1}{2},j} Y_{l_\kappa,\mu-\frac{1}{2}}(\bm{\hat{r}})\\
        C_{\mu+\frac{1}{2},-\frac{1}{2},\mu}^{l_\kappa,\frac{1}{2},j} Y_{l_\kappa,\mu+\frac{1}{2}}(\bm{\hat{r}})
    \end{pmatrix}
\end{eqnarray}
where $C$ are the Clebsch-Gordan coefficients. $\kappa$ and $\mu$ are angular momentum quantum numbers, with $\kappa$ taking positive and negative nonzero integer values, while $\mu$ takes half-integer values such that $-|\kappa| < \mu < |\kappa|$. Here the spherical bispinors satisfy the orthogonality property
\begin{eqnarray}
    \int_0^{2\pi}\int_0^\pi X_{\kappa',\mu'}(\theta,\phi)^\dagger X_{\kappa,\mu}(\theta,\phi) \sin \theta d\theta d\phi = \delta_{\kappa'\kappa}\delta_{\mu'\mu}.
\end{eqnarray}

\begin{figure*}[ht]
\centering
\includegraphics[width=10cm,height=8cm]{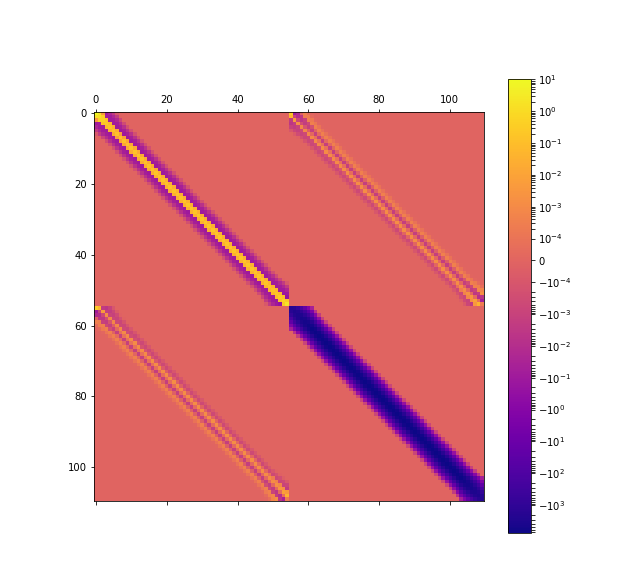}
\caption{\label{fig:radialband} Zoomed-in view of radial block matrix for $N=110$ 7th-order splines, chosen to make the band structure more clearly visible. Off-diagonal bands connect the positive- and negative energy components.}
\end{figure*}

For the radial functions, we use a modified B-spline basis implementing the so-called dual kinetic balance \cite{Shabaev2004}. B-splines \cite{H.Bachau_2001} are piecewise polynomials with compact support which have been shown to be highly accurate for approximating the radial components of atomic systems and see regular usage in solutions of the Dirac equation \cite{PhysRevA.37.307}. 
The B-splines are defined on a knot interval $t_0,t_1,...,t_N$ through a simple recursive formula starting with the $0$th-order splines, defined as
\begin{eqnarray}
    u_{k,0}(x) = \left\{\begin{matrix}
        1 & t_k < x \leq t_{k+1} \\
        0 & \text{otherwise.}
    \end{matrix}\right.
\end{eqnarray}
Then, the higher-order splines are given by the recursion formula \cite{DeBoorCarl1978Apgt}:
\begin{eqnarray}
    u_{k,d} = \frac{x-t_k}{t_{k+d}-t_k}u_{k,d-1}(x) + \frac{t_{k+1+d}-x}{t_{k+d+1}-t_{k+1}} u_{k+1,d-1}(x).
\end{eqnarray}
To obtain a set of basis functions that approximates arbitrary (smooth) functions in the knot interval, the knot interval may be "padded" by repeating the first and last knots $d$ times.
For this problem, the B-splines are implemented with a knot vector that is in principle arbitrary, but using an evenly-spaced grid padded at the boundaries by default.
The dual kinetic balance form resolves the issue of "spurious states" which arises when an inappropriate choice of basis is used to discretize the Dirac equation. The radial basis functions are given by:
\begin{eqnarray} \label{eqn:dualkinetic}
    \begin{pmatrix}
            P_{k,\kappa}(r) \\ 
            Q_{k,\kappa}(r)
        \end{pmatrix}&
        = &
        \begin{pmatrix}
            u_{k}(r) \\ 
            \frac{1}{2c} \left(\frac{d}{dr} + \frac{\kappa}{r}\right) u_{k}(r)
        \end{pmatrix}, k \leq N\\
        \begin{pmatrix}
            P_{k,\kappa}(r) \\ 
            Q_{k,\kappa}(r)
        \end{pmatrix}&
        = &
        \begin{pmatrix}
            \frac{1}{2c} \left(\frac{d}{dr} - \frac{\kappa}{r}\right)u_{k-N}(r) \\ 
            u_{k-N}(r)
        \end{pmatrix} , N < k \leq 2N.
\end{eqnarray}
The $P,Q$-doublets' mathematical properties also affect the matrix structure in a meaningful way. The B-splines are not orthogonal, however their compact support results in a useful constraint on the sparsity pattern of the matrix: Splines of degree $d$ only overlap with their $d$ nearest neighbors, meaning that for a given $(P,Q)$-doublet couplings occur only between nearest neighbors and modulo-$N$ nearest neighbors. This results in a distinct band structure as shown in Figure \ref{fig:radialband}, with the diagonal band corresponding to nearest-neighbor couplings, and off-diagonal bands corresponding to the coupling between $P$ and $Q$.

In our simulations, we consider a vector potential of the electromagnetic fields satisfying the Coulomb gauge condition, $\nabla \cdot \bm{A} = 0$. We use a carrier-envelope form of a linearly polarized pulse with a $\sin^2$ envelope and a sinusoidal carrier wave:
\begin{eqnarray}\label{eqn:ndpPot}
    \bm{A}(t,\bm{r}) = \\\nonumber
        \left\{ \begin{matrix}
        \frac{E_0}{\omega}\sin^2\left(\frac{\pi}{T}t - \frac{\pi}{cT}\bm r \cdot \bm{\hat{k}}\right) \sin \left(\omega t - \bm{k}\cdot \bm{r}\right) \bm{\hat{u}}_p, && 0 < t - \bm r \cdot \bm{\hat{k}} < T,\\
        0, && \text{otherwise.}
    \end{matrix}\right.
\end{eqnarray}
This pulse expression is used to model a finite-duration, spatially localized pulse propagating along the direction of k, with temporal duration $T$, carrier frequency $\omega$ and the pulse intensity $E_0$. The term $\sin^2$ acts as a traveling envelope that ensures the pulse is non-zero only during the time interval $0 < t - \bm r \cdot \bm{\hat{k}} < T$.

This form may appear non-standard, as it is physically equivalent to a pulse generated by a source emitting for a finite duration $T$. In our case, the choice of a $\sin^2$ envelope ensures smooth turn-on and turn-off of the field, minimizing unphysical high-frequency components. In addition, this form is advantageous as the time- and space-dependence may be separated using trigonometric identities, allowing for matrix elements to be calculated in a time-independent manner before the simulation run. Further, the potential may be expressed in spherical coordinates through plane wave expansion. This results in an expression for the spatial part of the potential as a sum over products of spherical harmonics and Bessel functions, while the time-dependence is factorized into scalar functions $f_\alpha(t)$:
\begin{eqnarray}\label{eqn:bdpA}
    \frac{E_0}{\omega}\bm A(t,\bm r) = 
    \bm{\hat{u}}_p\sum_{\alpha=1}^6 f_\alpha(t) \sum_{l=0}^{l_{max}} \sum_{m=-l}^{l} g_{\alpha,l}(r) Y_{lm}(\bm{\hat{k}})^*Y_{lm}(\bm{\hat{r}}).
\end{eqnarray}
Here, $\bm{\hat{u}}_p$ is the polarization vector, and $\bm{\hat{k}}$ is the propagation direction, while $r$ and $\bm{\hat{r}}$ are the magnitude and direction components of $\bm{r}$, respectively. The functions $f_\alpha(t)$ are sine and cosine functions arising from the application of trigonometric identities to a $\sin^2$ envelope function, while $g_{\alpha,l}(r)$ are Bessel functions corresponding to a plane wave expansion of the spatial part of the function.

By incorporating Eq.(\ref{eqn:basis}) into the TDDE Eq.(\ref{eqn:TDDE}), the matrix elements of the Dirac Hamiltonian take the following form
\begin{eqnarray}
    \bm{H}_{ij} = \int \begin{pmatrix}
        P_{k_i\kappa_i} X_{\kappa_i \mu_i}\\
        iQ_{k_i\kappa_i} X_{\kappa_i\mu_i}
    \end{pmatrix}^\dagger H \begin{pmatrix}
        P_{k_j\kappa_j} X_{\kappa_j \mu_j}\\
        iQ_{k_j\kappa_j} X_{\kappa_j\mu_j}
    \end{pmatrix}  \sin \theta  dr d \theta d\phi,
\end{eqnarray}
which through manipulation of the basis functions and exploitation of the linearity and radial separability of the Dirac equation, may be expressed in terms of smaller block matrices as
\begin{eqnarray}\label{eqn:fullH}
    (\bm{H})_{\kappa',\kappa,\mu'\mu}(t) = \\\nonumber
    (\bm{H}_0)_{\kappa'\kappa\mu'\mu} + i \sum_{\alpha = 1}^6 f_\alpha(t) \sum_{l=0}^{l_{max}} (\bm{H}^{\theta})_{l,\kappa',\kappa,\mu',\mu} \left[ \bm g_{\alpha,l}^0 + \kappa' \bm g_{\alpha,l}^1 + \kappa \bm g_{\alpha,l}^2 + \kappa'\kappa \bm g_{\alpha,l}^3\right]\\\nonumber
    - (\bm{H}^{\theta})_{l,\kappa,\kappa',\mu,\mu'} \left[ (\bm g_{\alpha,l}^0)^\dagger + \kappa (\bm g_{\alpha,l}^1)^\dagger + \kappa' (\bm g_{\alpha,l}^2)^\dagger + \kappa'\kappa (\bm g_{\alpha,l}^3)^\dagger\right]\\\nonumber
    (\bm{H}_0)_{\kappa'\kappa\mu'\mu} = \delta_{\kappa'\kappa}\delta_{\mu'\mu}\left(\bm{h}_{0} + \kappa\bm{h}_{1} + \kappa^2\bm{h}_{2} + \kappa^3\bm{h}_{3}\right).
\end{eqnarray}
Here, the matrices $\bm g_{\alpha,l}^n$ project the functions $g_{\alpha,l}$ onto the radial basis functions Eq. \eqref{eqn:dualkinetic}, while the matrices $\bm{H}_{l}^\theta$ contain the projection of the angular component of Eq. \eqref{eqn:bdpA} onto the spinor basis elements Eq. \eqref{eqn:spinorbasis}, and the matrices $\bm{h}_n$ correspond to the projection of the radial Hamiltonian on the basis.

Numerical integrals of products between the B-splines are evaluated through Gauss-Legendre quadratures, taking advantage of the splines being piecewise polynomials to set the number of quadrature points according to the degree of the splines. By discarding splines which are themselves nonzero at the $r=0$ boundary, or have derivatives which are nonzero, the zero-boundary condition for the radial component of the wavefunction is enforced. When sufficiently high-order splines are used we ensure that $\frac{u_k(r)}{r^n} \rightarrow 0$ as $r\rightarrow 0$ for all basis functions, such that the singularity is dealt with. Here, a degree of $d=7$ was found to be sufficient to satisfy the accuracy requirements. For the opposite boundary at $r=r_{\text{max}}$, boundary conditions may be enforced by a similar choice of splines. 

Details of how the Hamiltonian submatrices $\bm{g}_{\alpha,l}^n$, $\bm{h}_n$ and overlap submatrices $\bm{s}_n$ are calculated are given in \cite{Johanne2024}, and for convenience we give the definition of the coupling matrices $\bm{H}_l^\theta$ here:
\begin{eqnarray}\label{eqn:Htheta}
    (\bm{H}^{\theta}_l)_{\kappa',\kappa,\mu',\mu} = 4\pi Y_{l0}(\bm{\hat{z}})\left<X_{-\kappa',\mu'}
    \right|\sigma_x Y_{l0}(\bm{\hat{r}})\left|X_{\kappa,\mu} \right>.
\end{eqnarray}
With this scheme we may rapidly assemble the full time-dependent matrix at each time step with minimal redundant calculations necessary, in particular calculating $\bm{H}_0$ only once at the beginning of the simulation. This means that the numerical integration procedure only needs to be performed once, with the matrices assembled as needed according to Eq. \eqref{eqn:fullH} from the stored $\bm{h}$ and $\bm{g}$-matrices.

\subsection{Initial states and eigenvalue problems}
To obtain initial states of the system, we solve the eigenvalue problem
\begin{eqnarray}
    \bm{H}_0 \Psi(0) = \bm{S}\Psi(0).
\end{eqnarray}
However, in line with the partitioning scheme presented above, we take advantage of the repeating structure in $\bm{H}_0$ to skip large portions of the eigenvalue problem. By Eq. \ref{eqn:fullH}, any given block of $\bm{H}_0$ depends only on $\kappa$, with matrix elements for differing values of $\mu$ identical. Thus, by solving the smaller eigenvalue problem
\begin{eqnarray}
    \left(\bm{h}_{0} + \kappa\bm{h}_{1} + \kappa^2\bm{h}_{2} + \kappa^3\bm{h}_{3}\right) \Psi_\kappa = \left(\bm{s}_{0} + \kappa\bm{s}_{1} + \kappa^2\bm{s}_{2} \right)\Psi_\kappa,
\end{eqnarray}
we obtain all eigenstates for the given $\kappa$ simultaneously, skipping an increasing number of redundant calculations as the number of $\mu$-states grows for higher $|\kappa|$. The computational complexity of the problem is then reduced to solving $2\kappa_{\text{max}}$ separate eigenvalue problems with sparse matrices of dimensions $2N\times2N$, and the scaling of the problem depends primarily on $N$. As the time evolution is by far the more demanding task setting restrictions on grid size, scaling of the eigenvalue problem beyond sizes manageable for time evolution was not considered.

\subsection{Time propagation}

The time evolution of the system is formally performed using a matrix that satisfies

\begin{eqnarray}
    i \hbar \pd{}{t} \bm{U}(t,t_0) = \bm{S}^{-1}\bm{H}(t) \bm{U}(t,t_0).
\end{eqnarray}
This equation does not have an exact closed-form solution for time-dependent $\bm{H}$. However, in sufficiently short time intervals, an approximate solution can be
\begin{eqnarray}
    \bm{U}(t,t_0) \approx e^{- i \hbar \int_{t_0}^t \bm{S}^{-1} \bm{H}(t)} \approx e^{- i \hbar \bm{S}^{-1} \bm{H}\left(t_0 + \Delta t / 2\right) \Delta t},
\end{eqnarray}
where the time integral of the matrix is approximated by the midpoint rule, holding as long as $\Delta t$ is sufficiently small so that $\bm{H}$ is approximately constant with $\textsc{O}(\Delta t^2)$ error. The final state of the system is then retrieved by incremental time steps
\begin{eqnarray}
    \psi(t_N) = U(t_N,t_{N-1})U(t_{N-1},t_{N-2})\times ... \times U(t_2,t_1)U(t_1,t_0)\psi(t_0).
\end{eqnarray}
Implementing this procedure still requires exponentiating a very large matrix, which is an undesirable operation. As such, we make use of the first-order Padé approximant of the matrix exponential,
\begin{eqnarray}
    e^{\bm{B}}\approx \left(1 - \bm{B}/2\right)^{-1}\left(1 + \bm{B}/2\right),
\end{eqnarray}
resulting in the Crank-Nicholson form propagator
\begin{equation}\label{eqn:solPsi}
    \Psi\left(t + \Delta t\right) = \left(\bm S + i \frac{\hbar\Delta t}{2} \bm H\left(t+\Delta t\right)\right)^{-1}\left(\bm S - i \frac{\hbar\Delta t}{2} \bm H\left(t+\Delta t\right)\right)\Psi(t).
\end{equation}
This choice of approximation is advantageous due to how it works around the factor of $\bm{S}^{-1}$, retaining sparsity for both the numerator and denominator of the time evolution operator, and is a popular choice for B-spline methods in atomic physics \cite{ECormier1997}. As an implicit method, this yields comparatively stable time evolution at the cost of requiring the solution of an inverse linear system. While not ideal, this is still feasible using the iterative algorithm BiCGSTAB. In fact, the form of the problem already provides us with a useful preconditioner for the system: Due to its simple block-diagonal structure, we may use an LU decomposition of the matrix
\begin{eqnarray}\label{eqn:matrixM}
    \bm{M} = \bm{S} + i \frac{\Delta t}{2} \bm{H}_0,
\end{eqnarray}
which is easily performed at startup using standard linear algebra algorithms to obtain very sparse $\bm{L}$ and $\bm{U}$-matrices, efficiently preconditioning the system to ensure fast convergence.

The time evolution is now a problem of simply calculating matrix-vector products quickly. Due to its role in determining the sparsity pattern of $\bm{H}$, restructuring and partitioning based on the large-scale coupling matrix $\bm{H}^\theta$ is particularly advantageous. This can be expanded based on Eq. (\ref{eqn:Htheta}) to:
\begin{eqnarray}\label{eqn:HthetaExpand}
    H^{\theta}_{l,\kappa',\kappa,\mu',\mu} = C_{\mu'-\frac{1}{2}, \frac{1}{2},\mu'}^{l_{-\kappa'},\frac{1}{2},j'}
    C_{\mu-\frac{1}{2}, \frac{1}{2},\mu}^{l_\kappa,\frac{1}{2},j} 
    \left<Y_{l_{-\kappa'},\mu'-\frac{1}{2}}(\bm{\hat{r}})
    \right|Y_{l0}(\bm{\hat{r}})\left|
    Y_{l_\kappa,\mu+\frac{1}{2}}(\bm{\hat{r}}) \right> \\\nonumber 
    + C_{\mu'+\frac{1}{2},-\frac{1}{2},\mu'}^{l_{-\kappa'},\frac{1}{2},j'}
    C_{\mu+\frac{1}{2},-\frac{1}{2},\mu}^{l_\kappa,\frac{1}{2},j} 
    \left<Y_{l_{-\kappa'},\mu'+\frac{1}{2}}(\bm{\hat{r}})
    \right|Y_{l0}(\bm{\hat{r}})\left|
    Y_{l_\kappa,\mu-\frac{1}{2}}(\bm{\hat{r}}) \right>,
\end{eqnarray}
where the integral of the product of the spherical harmonic functions is evaluated in terms of the Wigner 3j-symbols
\begin{eqnarray}\nonumber
    \int_0^{2\pi}\int_0^\pi Y_{l_1 m_1}(\theta,\phi)Y_{l_2 m_2}(\theta,\phi)Y_{l_3 m_3}(\theta,\phi) \sin \theta d\theta d\phi\\\label{eqn:GauntCoef}
    =
    \sqrt{\frac{(2l_1 + 1)(2 l_2 + 1)(2 l_3 + 1)}{4\pi}}
    \begin{pmatrix}
        l_1 && l_2 && l_3\\
        0 && 0 && 0
    \end{pmatrix}
    \begin{pmatrix}
        l_1 && l_2 && l_3\\
        m_1 && m_2 && m_3
    \end{pmatrix}.
\end{eqnarray}

\begin{figure*}[ht]
\centering
\includegraphics[width=8cm,height=7cm]{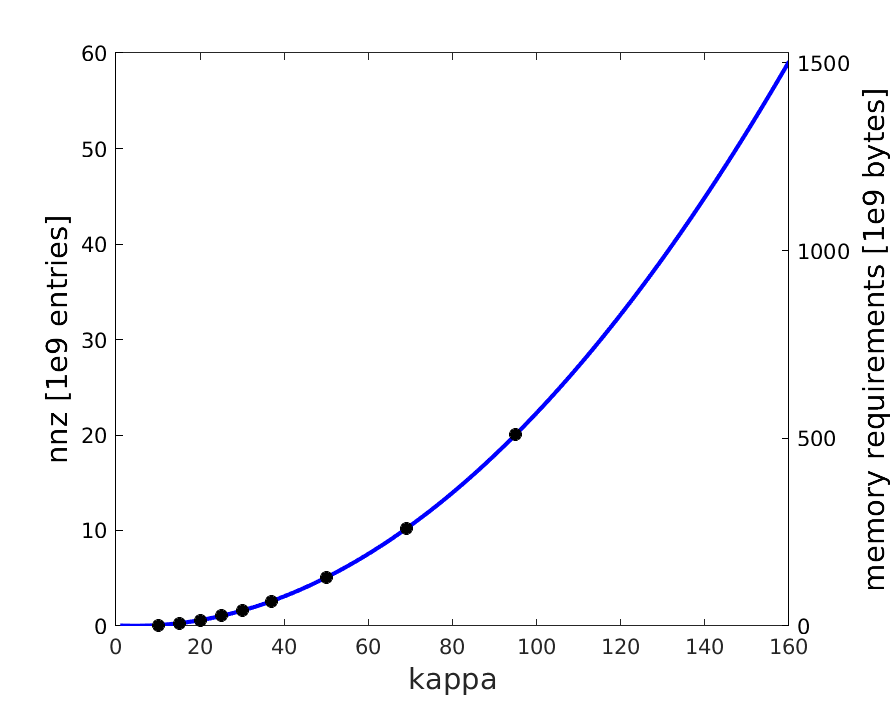}
\caption{\label{fig:kappa} The number of non-zero elements (nnz, corresponding to memory requirement) as a function of $\kappa$. Note that for each $\kappa$, the values of $\mu$ is determined by $-|\kappa| < \mu < |\kappa|$.}
\end{figure*}

Nonzero elements of $\bm{H}^\theta$ only appear for combinations of $\kappa',\kappa,\mu',\mu$ which satisfy the exclusion rules present in Eq. (\ref{eqn:HthetaExpand}). We illustrate the relationship between nonzero elements and $\kappa$ in Fig. \ref{fig:kappa} highlighting the required memory for a given $\kappa$.

The structure of $\bm{H}_\theta$ in the numerical calculations is determined by three parameters: $\kappa_{\text{max}}$, $\mu_{\text{max}}$ and $l_{\text{max}}$. The $l$ here should not be confused with the orbital angular momentum quantum number $l$, but instead refers to the $l$ in Eq. \eqref{eqn:HthetaExpand}. $\kappa_{\text{max}}$ and $\mu_{\text{max}}$ act to determine the spherical resolution of the problem, while $l_{\text{max}}$ determines the number of terms in the expansion of the nondipole field. The exclusion rules arising from Eq. \eqref{eqn:GauntCoef} for this matrix indicate that $\bm{H}^\theta$ will only have nonzero coefficients when
\begin{eqnarray}\\\nonumber
    \mu = \mu' \pm 1\\
    \left||-\kappa| - |\kappa'| + \frac{\sgn{-\kappa} - \sgn{\kappa'}}{2}\right| \leq l \leq |-\kappa| + |\kappa'| + \frac{\sgn{-\kappa} + \sgn{\kappa'}}{2} - 1.
\end{eqnarray}
Therefore, increasing $l_\text{max}$ increases the number of nonzero matrix elements as more combinations of $\kappa',\kappa$ become coupled by the nondipole terms, incurring a computational cost for higher values. There is however a convenient bound on what values of $l_\text{max}$ are sufficient: Within a finite-radius simulation domain the functions $g_{\alpha,l}(r)$ go to zero at higher values of $l$. For the pulse parameters used here, we find that $l_\text{max} = 10$ is sufficient.

\section{GaDE implementation}\label{computing}

\subsection{Architecture of GaDE}

\begin{figure*}[ht]
\centering
\includegraphics[width=12cm,height=10cm]{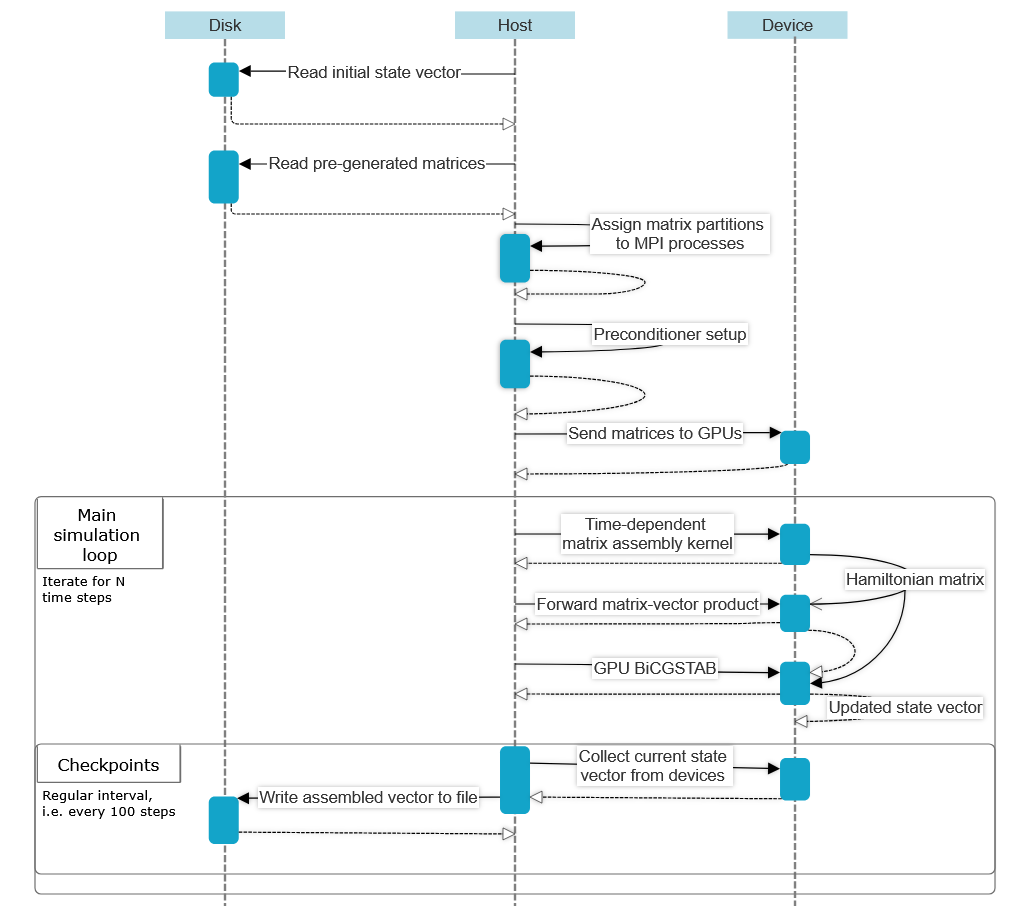}
\caption{\label{fig:diagram} UML diagram showing the high-level structure of the GaDE solver, and how tasks are distributed through the hardware: Disk activity is limited to loading pre-generated data and saving the state vector at checkpoints, while computationally intensive linear algebra computations are handled by the GPUs.}
\end{figure*}

The high-level architecture of the GaDE solver is presented in Fig. \ref{fig:diagram}. From the implementation perspective, a GaDE simulation consists of:

\begin{enumerate}
\item Pre-computation of the input data of GaDE.

  \begin{enumerate}
  \item Computation of sub-matrices $\bm g$ and $\bm h$ and the matrix $\bm H^\theta$ [cf. Eq. \eqref{eqn:fullH}].
  \item Computation of a METIS partitioning based on the $\bm H^\theta$ matrix.
  \item Computation of the stationary matrix $\bm{S}$ and the preconditioner $M$ [cf. Eq. \eqref{eqn:matrixM}].
  \end{enumerate}
  
\item Time propagation of TDDE: in each time step.
  \begin{enumerate}
  \item Computation of $\bm{H}(t+\Delta t)$,
  \item Using BiCGSTAB to obtain the solution of Eq. (\ref{eqn:solPsi}).
  \end{enumerate}
\end{enumerate}

\noindent The input data (1a) is pre-computed sequentially,
off-line, before the time-dependent part of the simulation. This choice is made to simplify
the computationally heavy part of the code, and because these calculations are not a performance
bottleneck. In addition, most of those input matrices can be re-used in different simulation scenarios.

Partitioning (distribution of $\bm{H}$ and the degrees of freedom among MPI ranks) of the system for parallel execution (1b) is also performed on a single CPU, using a python script. This step applies METIS to the relatively small matrix $\bm{H}^\theta$,
to renumber the degrees of freedom in the system. The goal is to maximize load balancing and minimize communication between MPI ranks (see section \ref{sec:parallel}).

The remaining computations (1c and 2) are performed fully in parallel using MPI. The GaDE solver can work both on CPUs, and on GPUs. In the latter case each MPI rank uses one dedicated GPU for
calculations. The detailed implementation steps of GaDE are:

\begin{enumerate}
\item each MPI rank reads the $\bm H^\theta$, $\bm g$ and $\bm h$ matrices, and the partitioning information from input files.
\item based on the partitioning, each rank pre-allocates the memory for the locally owned parts of
 system matrices $\bm{H}$ and $\bm{S}$ on the target compute resource used (either CPU, or GPU).
 This way memory allocation is only done once, which is crucial for performance reasons.
\item each rank analyzes the communication pattern of Sparse Matrix - Vector product (SpMV) $\bm{H} \cdot x$ and pre-allocates the communication buffers (see section \ref{sec:parallel}).
\item each rank allocates and computes the preconditioner $\bm{M}$ on the target compute resource.
\item if computing on GPUs, the input matrices ($\bm g$ and $\bm h$) are copied to the device once,
 before the time propagation loop. They are then used in each time step, during computation of $\bm{H}$.
\item time propagation loop.
\end{enumerate}

\subsection{Parallelization}\label{sec:parallel}

\begin{figure*}[ht]
\centering
\includegraphics[width=6.5cm,height=6cm]{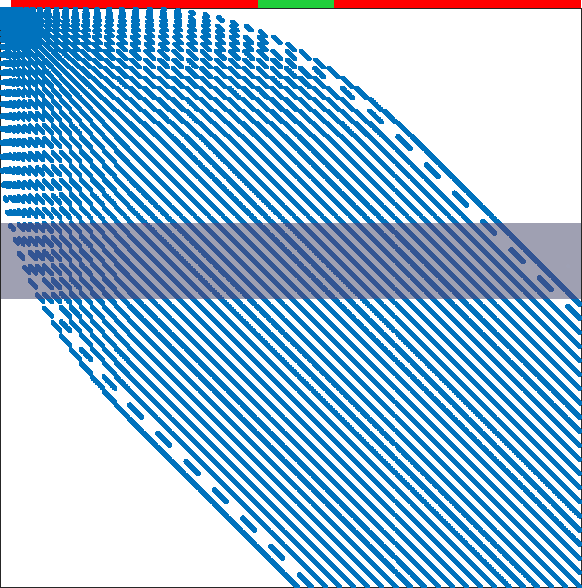}
\includegraphics[width=6.5cm,height=6cm]{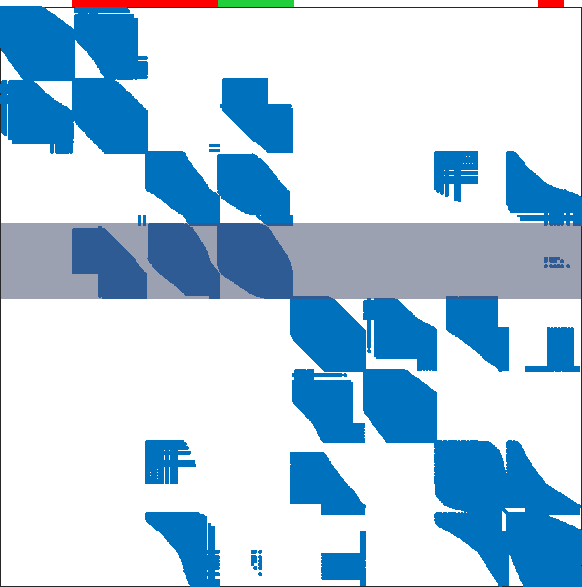}
\caption{\label{fig:sparse} Non-zero structure of $\bm{H}$: Original matrix (Left-hand side) and matrix partitioned between 8 ranks (right-hand side). Shaded part: matrix part owned by an example partition. Green (local) and red (non-local) vector entries referenced during SpMV.}
\end{figure*}

Efficient parallelization relies on dividing the problem
solved among the MPI ranks so that 1) the amount of work performed by each rank is similar, and 2) the amount
of communication between the ranks is minimized. Load imbalance and communication are among the main reasons why parallel codes might not scale; imbalance introduces idle time (all ranks wait for the slowest one), while communication is an overhead and not useful computation. In GaDE, computation of $\bm{H}$ entries and SpMV are
the most computationally demanding parts. In this context parallelization is achieved by
distributing matrix $\bm{H}$ and vector $x$ among the MPI ranks. Usually in parallel SpMV
this is done by assigning contiguous rows of the matrix to individual processes.

Distribution of the vector follows naturally: each rank
owns the vector entries updated during SpMV. Figure \ref{fig:sparse} shows an example matrix $\bm{H}^\theta$
and one submatrix owned by a chosen rank (shaded area).
Note that SpMV requires the input vector entries
that correspond to the non-zero columns in the local matrix part (span of the input vector marked by red
and green colors in the figure). In general, those vector entries
can be owned by other ranks (red vector parts) and thus result in communication: before each SpMV each MPI rank needs
to obtain the needed vector entries from their owners.

For spatial problems (e.g. 3-dimensional PDEs) a common approach is to use graph partitioning
algorithms \cite{Chevalier_2008,Karypis_1998}.
These libraries balance the local problem size (graph vertices, in SpMV - vector entries) and work (graph edges,
in SpMV - number of non-zeros in per-rank sparse matrix) with inter-partition connectivity that results
in communication (graph edges connecting different partitions, in SpMV - matrix entries
that reference non-local vector parts). METIS renumbers the degrees of freedom in the system in a way
that groups the densely connected sub-graph vertices. The rows and columns of the system matrix must be
permuted accordingly. METIS is a well established method of graph partitioning that maximizes load balancing, while at the same time minimizing the communication (edge cut between partitions and the number of partition neighbors). METIS and other similar graph partitioning codes are a de facto standard when partitioning mesh-based problems, with their effectiveness well studied in the literature, e.g. \cite{Gropp2000}. The importance of problem decomposition in the context of Sparse Matrix - Vector multiplication (SpMV) is well demonstrated in \cite{Bienz2019} and references within. In short, the fact that the reduction of communication and load imbalance, both of which are overheads that slow down computations, improves performance is a well established consensus and one of the main targets for optimization in this type of problems.

This approach yields very good results for the TDDE discretization discussed in this paper.
Figure~\ref{fig:sparse} shows the non-zero structure of the original coupling matrix $\bm{H}^\theta$ (Fig.~\ref{fig:sparse}~(left))
and the reordered matrix, which is suitable for parallel calculations on 8 ranks (Fig.~\ref{fig:sparse} (right)).
Shaded matrix part is assigned to one specific MPI rank.
The red areas at the top of the matrix show the non-local vector
entries referenced by the matrix part during SpMV. The green areas denote the local vector parts. 
Clearly, METIS substantially reduces the communication:
each of the partitions is only connected to 2-3 other partitions, which limits the number of MPI messages
that need to be sent. Moreover, the total volume of communication (number of exchanged vector entries) is smaller.

Note that implementing parallel SpMV always requires some approach to partition the matrix among ranks. In our particular case, we have demonstrated in Figure 4 that METIS provides an improvement over a naive partitioning that arises from direct discretization of the governing equations. We do not claim that METIS is the best approach to partition the problem, and we do not aim to rigorously compare it to any other approach - this is beyond the scope of this study. What we do show is that our current approach scales on thousands of GPUs with near perfect efficiency.
\subsection{GPU implementation}

When GaDE is using GPUs the time propagation loop is performed fully on the device.
Computation of $\bm{H}$ entries and sparse matrix assembly are implemented in a dedicated CUDA / HIP kernel.
Each non-zero entry in $\bm{H}$ is computed by one GPU thread. Blocks of threads compute subsequent
entries in the same row of the matrix. Since the sparse matrix is stored in CSR format,
writing the matrix entries to memory results in contiguous memory access and is efficient.

The BiCGSTAB implementation uses $cublas+cusparse$ on NVidia GPUs, and $hipblas+rocsparse$ on AMD GPUs.
The blas libraries are used for vector operations, $sparse$ libraries are used for an optimized SpMV implementation.
The code is mostly vendor-agnostic: HIP and CUDA APIs are very similar and the BiCGSTAB code is implemented
once, using function renaming depending on GPU type used. The only difference between HIP and CUDA implementations is how the SpMV is computed. In CUDA we use the $cusparse$
library. HIP has a similar library with identical APIs ($hipsparse$), which we could use with the
standard function renaming. However, we found that $rocsparse$ - a low-level implementation for AMD GPUs -
is approximately two times faster for our problem. Consequently, SpMV calls in GaDE use specialized functions
for both GPU architectures. 

HIP and CUDA only provide sequential implementations of SpMV, while parallel SpMV requires exchange of vector entries among the ranks. GaDE provides two implementations of the communication step, both of which require a GPU-aware MPI library \footnote{A GPU-aware MPI library allows calling the \textit{Send/Recv} routines directly on device pointers, without first copying the data to the CPU.}. In the first approach explicit packing kernels are used to copy the vector entries that need to be sent into a contiguous memory buffer located on the GPU. We then call \texttt{MPI\_Send} on the GPU buffer address. In the second approach we use MPI indexed types: the GPU vector pointer and indices of vector entries that need to be sent are passed to the \texttt{MPI\_Send} call. In this method it is the MPI library's responsibility to collect the non-contiguous data to be sent (GaDE does not run any packing kernels). The efficiency of those methods depends on the MPI library and the hardware, hence it is best to benchmark and use the better option.

\section{Performance and scalability analysis}\label{result}

In this section, we report on the performance of the GaDE solver by measuring the elapsed wall time, peak memory bandwidth, speed-up and parallel efficiency for both strong scaling and weak scaling. In our tests, floating-point operations are of double-precision complex type. Initially, we evaluate the solver's performance on a single GPU by considering different GPU architectures: NVIDIA A100, NVIDIA GH200 and AMD MI250X. The analysis is further extended to multiple GPUs across multiple nodes. In the latter experiment, the benchmarking data is generated from the supercomputer LUMI \cite{Lumi,LumiGuide}, specifically the GPU partition named LUMI-G, which is powered by AMD MI250X GPUs. In LUMI-G, each compute node consists of 4 AMD MI250X GPUs, where each AMD MI250X GPU features two separated Graphics Compute Dies (GCDs) with 64 GiB of high-bandwidth memory (HBM) per GCD. Two GCDs on the same AMD MI250X GPU are connected through Infinity Fabric with a bidirectional bandwidth of 400 GB/s, while the bandwidth is 200 GB/s for two GCDs located on two separate AMD MI250x GPUs (cf.~Table~\ref{tab:interconnect}). In addition, the CPU-to-GCD communication path provides a bidirectional bandwidth of 72 GB/s. On the other hand, each compute node comprises 4 HPE Cray Slingshot-11 network interface cards (NICs) - one for each MI250X GPU. Here, two GCDs share a NIC, and the communication path between two nodes through the HPE Slingshot interconnect provides a bidirectional bandwidth of 50 GB/s per NIC. We take advantage of the GPU-aware MPI: the inter-node communication follows an optimal path that connects first the GPU memory to the closest NIC without involving the CPU. This NIC end point, in turn, is connected to the other compute node's NIC through the HPE Slingshot interconnect. As a result, the additional communication overheads are minimized, thus reducing latency and leading to a significant improvement in the solver's performance.

For reference, we consider a CPU-based cluster for the MPI experiment. The CPU-partition is equipped with 2xAMD CPUs EPYC 7742, in which each compute node has 128 cpu-core. 

In our experiment on LUMI-G, the code is compiled using GCC 13.2.1 compiler, ROCm 6.0.3 and cray-mpich 8.1.29 libraries, and with GPU-aware MPI enabled.

\begin{table}[h!]
\centering
\begin{tabular}{|c|c|}
  \hline
  & Bidirectional \\
  Interconnect & bandwidth (GB/s) \\
  \hline
   \hline
  CPU-to-GPU & 72 \\
  GCD-to-GCD on same GPU & 400 \\
  GCD-to-GCD on different GPU & 100\\
  node-to-node per NIC & 50 \\
  \hline
\end{tabular}
\caption{Theoretical peak bidirectional bandwidths of various communication interconnects on the supercomputer \textit{LUMI-G}. The values are similar to those reported for the supercomputer \textit{Frontier} \cite{Yeung2025}.}
\label{tab:interconnect}
\end{table}

\subsection{Single GPU performance}

\begin{figure*}[ht]
\centering
\includegraphics[width=6.5cm,height=6cm]{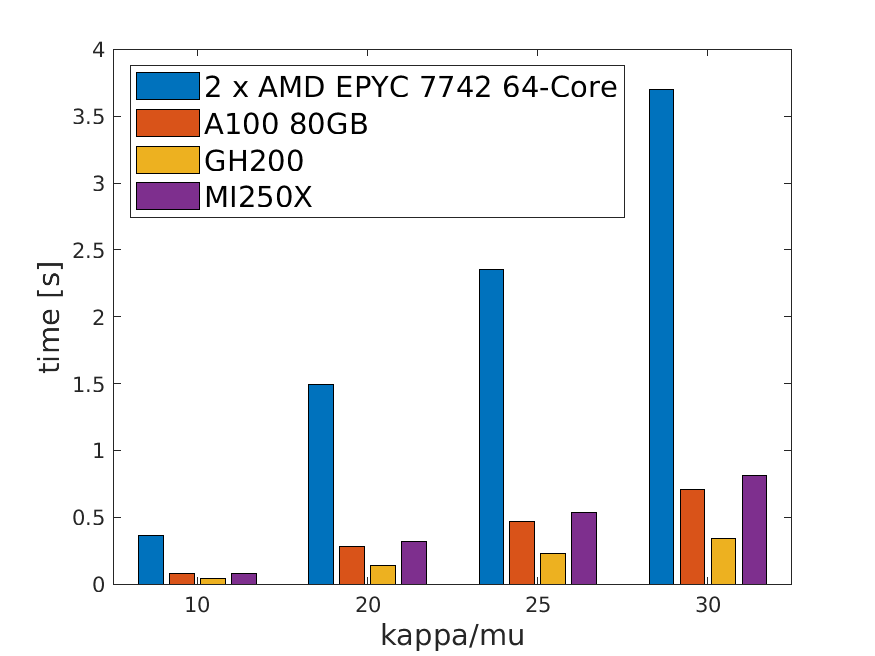}
\includegraphics[width=6.5cm,height=6cm]{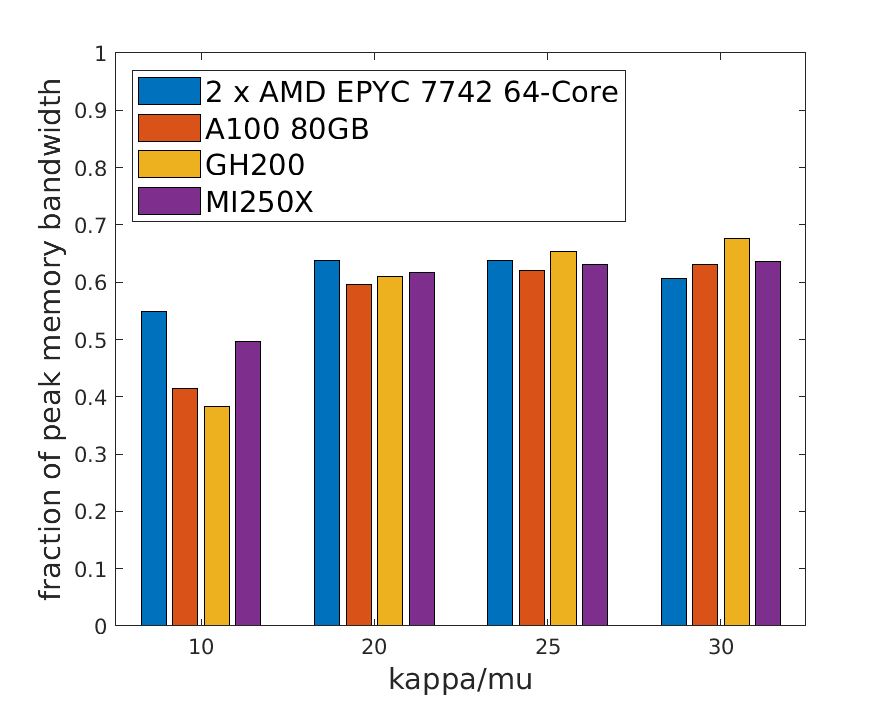}
\caption{\label{fig3} Single GPU performance comparison between different GPU architectures: NVIDIA A100 GPU, NVIDIA GH200 GPU and AMD MI250X GCD. The results from the 2-socket AMD EPYC 7742 system are also displayed. (Left-hand side) Elapsed wall time per step (s) as a function of $\kappa$. (Right-hand side) Fraction of peak memory bandwidths as a function of $\kappa$.}
\end{figure*}

Our initial experiment of the GaDE solver is summarized in Fig.~ \ref{fig3}. In particular, the execution time per time-step as a function of the problem size defined by $\kappa$ is shown in Fig. \ref{fig3}~(left) for the architectures mentioned above (cf.~Table~\ref{tab:gpu-arch}). For comparison, the performance of the CPU-based MPI implementation on a full compute node 2xAMD CPUs EPYC 7742 (128 cpu-core) is also shown. At first glance, the GPU implementation appears to outperform the CPU implementation. In particular, at $\kappa=30$, which corresponds to $1.6\times10^{9}$ non-zero elements, the solver's performance on the GPU is improved by a factor of 5 to 10 compared to the CPU version. On the other hand, one can see that the solver's performance on the NVIDIA GH200 GPU is nearly twice as fast as on A100 GPU and MI250X GPU. This discrepancy is related to the peak memory bandwidth of each architecture (cf.~Table~\ref{tab:gpu-arch}). This can be seen in Fig.~\ref{fig3} (right), which shows the measured fraction of the peak memory bandwidth. The results show a rate of about 65\% across all architectures. This demonstrates that the efficiency of the code is similar on all of the tested hardware, and that the performance of the GaDE solver is limited by the memory bandwidth.

\begin{table}[h!]
\centering
\begin{tabular}{|c|c|c|c|}
  \hline
                   & NVIDIA  & NVIDIA  & AMD \\
  GPU architecture & A100 GPU & GH200 GPU & MI250X GCD \\
  \hline
  \hline
  GPU Memory (GB) & 80 & 120 & 64  \\
  Peak memory bandwidth (GB/s) & 1935 &  4900 & 1630 \\
  CPU-GPU bandwidth (GB/s) & 32 & 450 & 36 \\
  Peak performance FP64 (TFLOPS) & 9.7 & 52 & 23.95\\
  Compiler & GCC 13 & GCC 13 & GCC 13 \\
  \hline
\end{tabular}
\caption{Specification of different GPU architectures: NVIDIA A100, NVIDIA GH200 and AMD MI250X.}
\label{tab:gpu-arch}
\end{table}

In the following, we further analyze the solver efficiency on multiple GPUs in a distributed environment. Here, we perform our analysis on the supercomputer LUMI-G. This is primarily due to its supercomputing capacity, allowing for large-scale of the solver's performance on thousands of GPUs. 

\subsection{Multiple GPU performance and scalability}

We examine the solver's performance and scalability on multiple GPUs across multiple compute nodes by analyzing strong scaling and weak scaling. This helps assess the efficiency and effectiveness of our solver and ensure optimal use of computational resources at large scale. The performance is evaluated by measuring the speedup and parallel efficiency for various problem sizes, while varying the added resources (number GPUs). These metrics are calculated using the following conventional formulas, respectively, for speedup 
\begin{eqnarray}\label{eqn:Speedup}
S = \frac{T_{single}}{T_{multi}}
\end{eqnarray}
and parallel efficiency 
\begin{eqnarray}\label{eqn:Efficiency}
P_e = \frac{1}{N_{p}} \frac{T_{single}}{T_{multi}} \,.
\end{eqnarray}
Here $T_{multi}$ is the execution time on multiple GPUs, and $T_{single}$ is the reference execution time on a single GPU (single GCD on MI250X AMD). $N_p$ refers to the number of GPUs.

\subsubsection{Strong scaling analysis}

\begin{figure*}[ht]
\centering
\includegraphics[width=6.5cm,height=6cm]{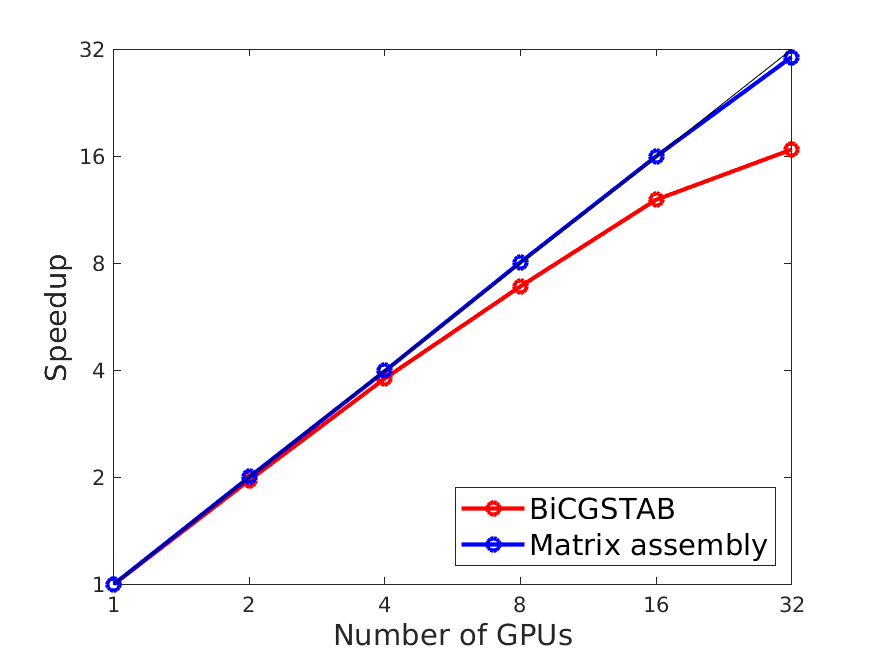}
\includegraphics[width=6.5cm,height=6cm]{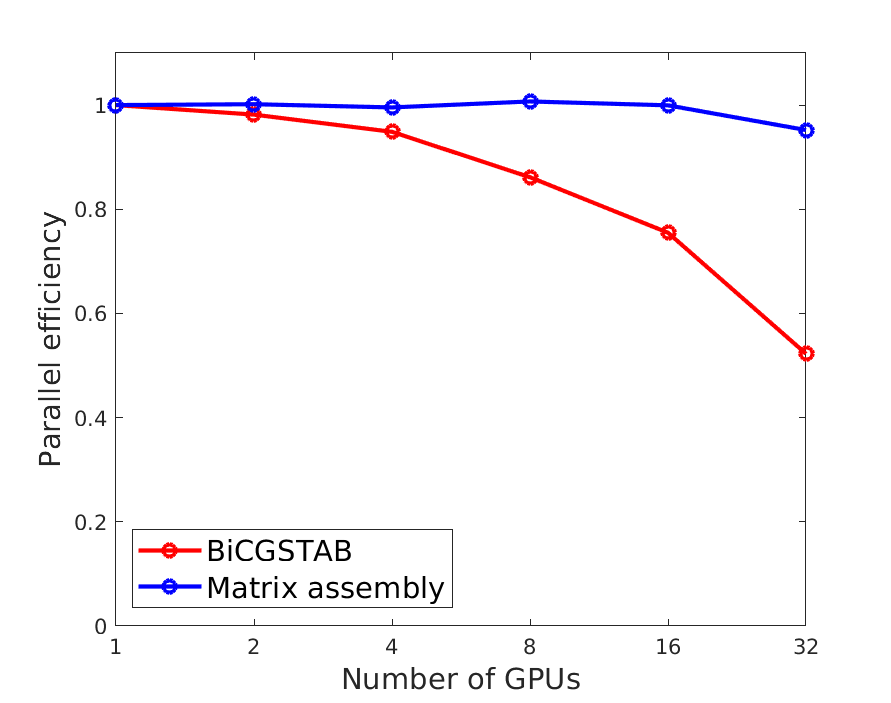}
\caption{\label{fig:strong-scaling} Strong scaling analysis of the \textit{Matrix assembly} (blue color) and \textit{BiCGSTAB} (red color). (Left-hand side) Speedup as a function of the number of GPUs. (Right-hand side) Parallel efficiency as a function of the number of GPUs.}
\end{figure*}

We start by evaluating the strong scaling scenario, in which the problem size is fixed at $\kappa=33$ (number of non-zero entries is $nnz$=$2\times10^9$, matrix dimension is $n$=$1\times10^6$), and the number of GPUs varies up to 32 GPUs. The results are shown in Fig.~\ref{fig:strong-scaling}. We measure the speedup (cf.~Fig.~\ref{fig:strong-scaling}~(left)) and parallel efficiency (Fig.~\ref{fig:strong-scaling}~(right)) for both \textit{Matrix Assembly} and \textit{BiCGSTAB}, which are the main components of the solver. The matrix assembly step does not perform any communication and as such it parallelizes perfectly. This is reflected in the linear speedup and a parallel efficiency of $\sim$ 1.

On the other hand, the speedup of \textit{BiCGSTAB} deviates from this linear scale when increasing the number of GPUs (and hence decreasing the problem size per-GPU), due to the growing communication cost relative to the computations. This is reflected in the picture of parallel efficiency where only 50\% of the performance is achieved at 32 GPUs. The efficiency curve indicates what per-GPU problem size should be used in large-scale runs and puts bounds on our weak scaling analysis.   

\subsubsection{Weak scaling analysis}

\begin{figure*}[ht]
\centering
\includegraphics[width=10cm,height=8cm]{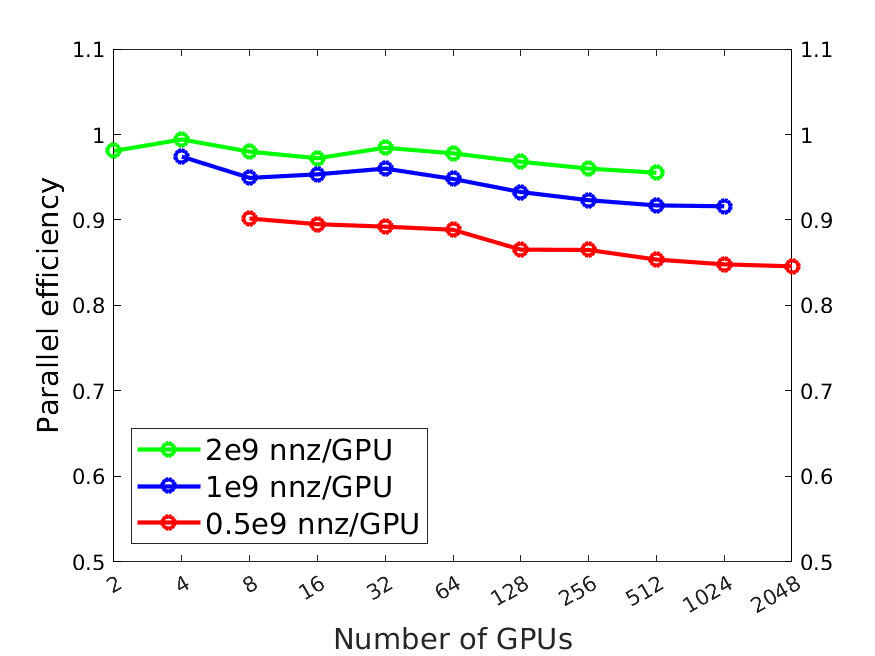}
\caption{\label{fig:weak-scaling} Weak scalability of \textit{BiCGSTAB} for three different problem sizes defined by the non-zero elements per GPU: $2\times10^9$ (green color), $1\times10^9$ (blue color) and $0.5\times10^9$ (red color).}
\end{figure*}

We extend our analysis of the GaDE solver to include weak scaling experiments. In general, weak scaling analysis provides insights into how well the solver maintains performance as the problem size grows proportionally to the number of computing resources. In our experiment presented in Fig.~\ref{fig:weak-scaling}, we measure parallel efficiency at different per-GPU problem sizes $nnz$: $0.5\times10^{9}$, $1\times10^{9}$ and $1\times10^{9}$ on 2048, 1024 and 512 GPUs, respectively. We focus our analysis on the \textit{BiCGSTAB} component, which is the computationally time-consuming part of the solver. As shown in Fig.~\ref{fig:weak-scaling}, the observed parallel efficiency remains nearly constant as the number of GPUs increases. Here we achieve excellent scalability with 85\% efficiency on 2048 GPUs, 92\% on 1024 GPUs and up to 96\% efficiency on 512 GPUs. 

In our tests, the per-GPU problem size is the number of non-zero entries in the local sparse matrix parts ($nnz$). This  corresponds to the amount of computational work that the GPU has to do during each time step. However, in our case this does not correspond equally well to the amount of communication: due to the quadratic scaling of $nnz$ with respect to $\kappa$ (Fig.~\ref{fig:kappa}) and a linear scaling of $n$ with respect to $\kappa$ (matrix dimension) it follows that the amount of per-GPU communication grows with total system size. This is one of the reasons for the parallel efficiency decline in Fig.~\ref{fig:weak-scaling}. In addition, increasing the problem size requires an increase of the number of MPI processes. This in turn might introduce additional overhead at large scale related to MPI communication (e.g., larger cost of collectives) or the fabric.

\subsection{Physical validation}
To validate convergence and physical correctness of the simulations for photoionization problems, a variety of methods were employed.

First, at lower intensities $(E_0=10-500)$, direct comparisons with dipole, nonrelativistic simulations were performed \cite{Johanne2024}. To explain deviations at higher pulse intensities, a thorough examination of the influence of both relativistic and nondipole corrections on the photoelectron momentum distributions was necessary \cite{Johanne2024.2}, with the observed shifts away from the nonrelativistic model found to be explainable through physical effects.

To demonstrate convergence in the high-intensity regime, as well as the practical performance of the GaDE solver, photoionization simulations were performed at high spatial and temporal resolutions using the hydrogen ground state for the initial vector, and the $\sin^2$ envelope pulse \eqref{eqn:bdpA} as the incident laser pulse at intensities up to $E_0=1000$ for the frequency $\omega=50$, and intensities up to $E_0=45$ for the frequency $\omega=3.5$. 

As lower frequencies lead to longer pulses, the $\omega=3.5$ case demanded a significantly larger simulation domain than the $\omega=50$ case despite the lower pulse intensity. In terms of grid resolutions, the $\omega=3.5$ case still proved less demanding than the highest pulse intensities for $\omega=50$ however.

As increasing simulation resolution demands a basis comprising higher-order spherical spinors, we use the differential photoelectron energy spectrum as our convergence metric. This observable is calculated as:
\begin{eqnarray}\label{eqn:dPdE}
    \frac{dP}{dE} = \sum_{\kappa,\mu} |\left<\psi_{\kappa,\mu,E}|\Psi\right>|^2.
\end{eqnarray}
Here, $\psi_{\kappa,\mu,E}$ is an eigenstate found through diagonalization of the Hamiltonian, for details see \cite{Johanne2024}.

Then, the error for a spectrum $P_n(E)$ may be defined relative to a given reference spectrum $P_m(E)$ as
\begin{eqnarray}
    \text{Err}_{n,m} = \int_{E_1}^{E_2}\left|P_n(E) - P_m(E)\right|^2 dE.
\end{eqnarray}

\begin{figure}
    \centering
    \includegraphics[width=\linewidth]{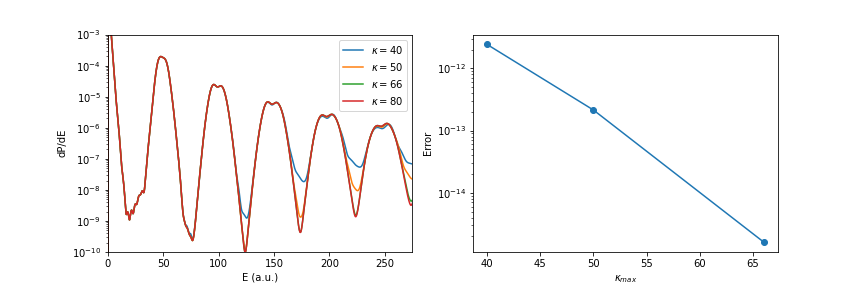}
    \caption{Convergence of GaDE results with increasing spatial and temporal resolutions, for physical parameters $\omega=50,3.5$ (top and middle, bottom) and $E_0=600,1000,45$. Increasing the angular resolution by adding more $\kappa,\mu$-states results in better convergence particularly at higher-energy parts of the spectrum.}
    \label{fig:convergence}
\end{figure}

The largest simulations in the convergence test data set were performed for a 15-cycle $\omega=50$ pulse in a spatial domain of $r_{\text{max}} = 60$, with B-splines defined on a radial grid of $1205$ knots and taking angular momentum states $|\kappa| \leq 80,|\mu| < 80$, running for $N=80000$ time steps, with pulse intensities of both $E_0 = 600$ and $E_0 = 1000$. One step down in resolution, simulations were also performed at $|\kappa| \leq 66,|\mu| < 66$ for $N=64000$ time steps at both intensities, with the resulting energy spectra shown in Fig. \ref{fig:convergence}.

For the $E_0=1000$ case, simulations performed at even resolutions were available for convergence testing, with $\kappa_\text{max} = 50$ and $40$, at a time resolution of $N=48000$ time steps. The relative error of these cases compared to the highest-resolution case is shown in Fig.  \ref{fig:convergence}, demonstrating a reduction in error with increased resolution. Note as well that low-energy parts of the spectrum converge faster than higher-energy parts, advantageous when primarily investigating one- or two-photon processes.

\begin{figure}
    \centering
    \includegraphics[width=\linewidth]{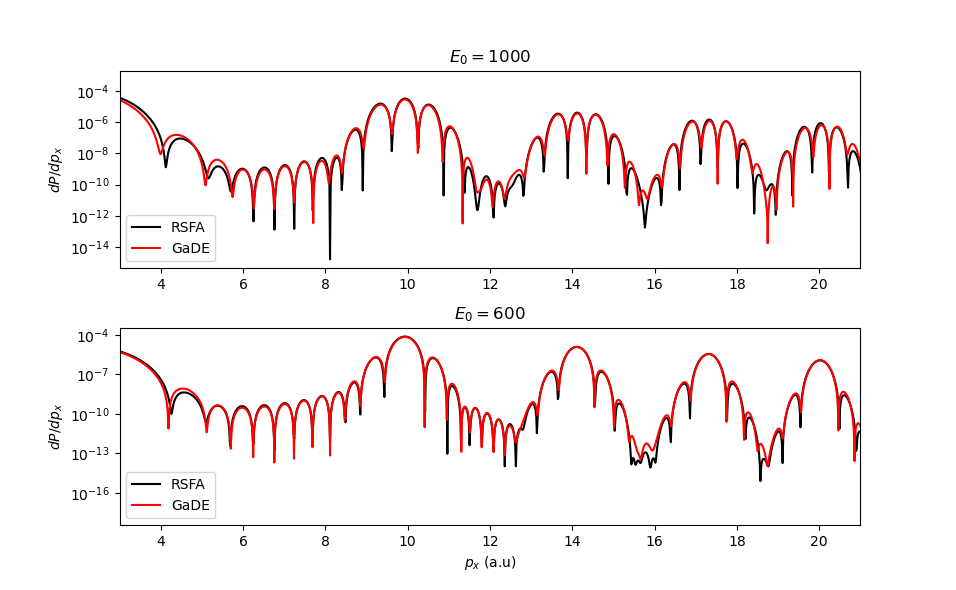}
    \caption{Comparison of GaDE solver results and relativistic strong-field approximation results finds qualitative agreement for both $E_0=1000$ and $E_0=600$ cases.}
    \label{fig:gadevsfa}
\end{figure}

For a second verification of the simulated results, the simulated photoelectron momentum distributions were compared with results obtained from a relativistic strong-field approximation \cite{Reiss1990} (rSFA) model. In this model, the photoelectron momentum distribution is obtained by
\begin{eqnarray}
    \Psi(\bm{p}) = \left<\psi^V\left|\gamma_\mu A^\mu\right|\phi_0\right>
\end{eqnarray}
Where $\phi_0$ is the initial state, here chosen to be the ground state, while $\psi^V$ is a relativistic \textit{Volkov state}, given by
\begin{eqnarray}
    \psi^V(x,p) = \frac{1}{(2 \pi)^{2/3} \sqrt{E_p}} \left(1 + \frac{1}{2 p \cdot k} k_{\mu} \gamma^\mu A_\nu \gamma^\nu\right) u \\\nonumber 
    \times \exp\left[-i p \cdot x - \frac{i}{p \cdot k} \int_{\eta}^{\infty} d \eta' \left( A \cdot p + \frac{1}{2}A^2\right) \right].
\end{eqnarray}
The Volkov state is a solution to the Dirac equation negecting the binding potential, and satisfies
\begin{eqnarray}
    (i\gamma^\mu \partial_\mu - \gamma^\mu A_\mu - c)\psi^V = 0.
\end{eqnarray}
As the incident laser pulse is very strong in the $\omega=50$ cases considered, the rSFA is expected to yield similar, but not identical results to the GaDE solver. Figure \ref{fig:gadevsfa} shows a comparison of results for the momentum distribution along the propagation direction for both $E_0=1000$ and $E_0=600$, finding good qualitative agreement between simulations and rSFA model. An error analysis was not performed, as highly precise quantitative agreement is not expected between this model and the full simulations.

\section{Conclusions}\label{conclusion} 
In conclusion, we have developed a GPU-enabled solver named GaDE to simulate the 3D time-dependent Dirac equation. The GaDE solver is designed to take advantage of the state-of-the-art supercomputing technologies that combines advanced GPU architectures and fast network interconnects with high-bandwidth and low latency.
We have discussed our implementation strategy that aims at performing almost the entire computation effectively on the GPUs. The implementation combines the use of the Message Passing Interface (MPI) library and CUDA/HIP programming models and leverages the GPU-aware MPI feature to optimize communication performance. This combination makes the solver portable across both NVIDIA GPUs and AMD GPUs in a distributed environment. Our experiments were conducted on the supercomputer LUMI, which is considered one of the world's top 10 fastest supercomputers. The supercomputer is equipped with AMD MI250X GPUs and HPE Slingshot interconnect. The GPU-related metrics were evaluated with a special focus on GPU memory bandwidth utilization, strong scaling, and weak scaling. Overall, we have achieved excellent scalability of up to 2048 GPUs with a parallel efficiency of 85\%. In addition, we have shown an optimal use of the GPU memory bandwidth achieving 65\% of the available capacity, although not reaching full utilization. Our analysis therefore demonstrates the effectiveness of the GaDE solver for large-scale computations.

In future research, the GaDE solver may be extended further to model many-body systems interacting with electromagnetic fields, which can be described in the picture of the time-dependent density functional theory. While such an extension would require significant coding work to implement the relativistic TDDFT Hamiltonian in place of the Dirac Hamiltonian, this modification has potential for several benefits to solid-state physics. Therefore, by leveraging modern heterogeneous HPC systems, the solver will enable users to uncover new physics in the relativistic regime, thus adding new insights into attosecond physics and strong-field physics and pave the way for exploring zeptosecond physics. 

\section*{CRediT authorship contribution statement}
\textbf{Johanne Elise Vembe:} Software, Methodology, Writing- Reviewing and Editing. \textbf{Marcin
Krotkiewski:} Software, Formal analysis, Methodology, Writing-
Reviewing and Editing. \textbf{Magnar Bjørgve:} Software, Reviewing. \textbf{Morten Førre:} 
Writing- Reviewing and Editing, Supervision. \textbf{Hicham Agueny:} Writing- Original draft,
Supervision, Methodology, Conceptualization, Project administration.

\section*{Code Availability} 
The source code developed in this work is publicly available under the AGPL-3.0 license at the GitHub Repo \cite{GitHub}.

\section*{Data availability} 
Data will be made available on request.

\section*{Declaration of competing interest} 
The authors declare that they have no known competing financial interests or personal relationships that could have appeared to influence the work reported in this paper.

\section*{Acknowledgments}
The MPI-based computations were performed on resources provided by Sigma2 - the National Infrastructure for High-Performance Computing and Data Storage in Norway. We acknowledge NRIS (Norwegian Research Infrastructure Services), Norway for awarding this project access to the LUMI supercomputer, owned by the EuroHPC Joint Undertaking, hosted by CSC (Finland) and the LUMI consortium through NRIS, Norway.

\bibliographystyle{elsarticle-num}
\bibliography{main_bib}

@article{Goulielmakis2010,
  title = {Real-time observation of
valence electron motion},
  author = {Goulielmakis, Eleftherios 
            and Loh, Zhi-Heng 
            and Wirth, Adrian 
            and Santra, Robin
            and Rohringer, Nina
            and Yakovlev, Vladislav S.
            and Zherebtsov, Sergey
            and Pfeifer, Thomas
            and Azzeer, Abdallah M.
            and Kling, Matthias F.
            and Leone, Stephen R.
            and Krausz, Ferenc},
  journal = {Nature},
  volume = {466},
  issue = {},
  pages = {739--743},
  numpages = {},
  year = {2010},
  month = {},
  publisher = {Nature},
  doi = {10.1038/nature09212},
  url = {https://doi.org/10.1038/nature09212}
}

@article{Sugioka2014,
  title = {Ultrafast lasers - reliable tools for advanced materials processing},
  author = {Sugioka, K. and Cheng, Y.},
  journal = {Light Sci Appl},
  volume = {3},
  issue = {},
  pages = {e149--e149},
  numpages = {4},
  year = {2014},
  month = {},
  publisher = {Nature},
  doi = {10.1038/lsa.2014.30},
  url = {https://doi.org/10.1038/lsa.2014.30}
}

@article{Johanne2024,
  title = {Relativistic and nondipole effects in multiphoton ionization of hydrogen by a high-intensity x-ray laser pulse},
  author = {Vembe, Johanne Elise and Johnsen, Esther A. B. and F\o{}rre, Morten},
  journal = {Phys. Rev. A},
  volume = {109},
  issue = {1},
  pages = {013107},
  numpages = {13},
  year = {2024},
  month = {Jan},
  publisher = {American Physical Society},
  doi = {10.1103/PhysRevA.109.013107},
  url = {https://link.aps.org/doi/10.1103/PhysRevA.109.013107}
}

@article{Agueny2018,
  title = {High-order photoelectron holography in the midinfrared-wavelength regime},
  author = {Agueny, Hicham and Hansen, Jan Petter},
  journal = {Phys. Rev. A},
  volume = {98},
  issue = {2},
  pages = {023414},
  numpages = {9},
  year = {2018},
  month = {Aug},
  publisher = {American Physical Society},
  doi = {10.1103/PhysRevA.98.023414},
  url = {https://link.aps.org/doi/10.1103/PhysRevA.98.023414}
}

@article{Chovancova2017,
  title = {Classical and quantum-mechanical scaling of ionization from excited hydrogen atoms in single-cycle THz pulses},
  author = {Chovancova, M. and Agueny, H. and R\o{}rstad, J. J. and Hansen, J. P.},
  journal = {Phys. Rev. A},
  volume = {96},
  issue = {2},
  pages = {023423},
  numpages = {10},
  year = {2017},
  month = {Aug},
  publisher = {American Physical Society},
  doi = {10.1103/PhysRevA.96.023423},
  url = {https://link.aps.org/doi/10.1103/PhysRevA.96.023423}
}

@article{Stacey1982,
  title = {Eliminating lattice fermion doubling},
  author = {Stacey, Richard},
  journal = {Phys. Rev. D},
  volume = {26},
  issue = {2},
  pages = {468--472},
  numpages = {0},
  year = {1982},
  month = {Jul},
  publisher = {American Physical Society},
  doi = {10.1103/PhysRevD.26.468},
  url = {https://link.aps.org/doi/10.1103/PhysRevD.26.468}
}

@article{Muller1998,
title = {Finite element formulation of the Dirac equation and the problem of fermion doubling},
journal = {Physics Letters A},
volume = {242},
number = {4},
pages = {245-250},
year = {1998},
issn = {0375-9601},
doi = {https://doi.org/10.1016/S0375-9601(98)00218-7},
url = {https://www.sciencedirect.com/science/article/pii/S0375960198002187},
author = {C. Müller and N. Grün and W. Scheid}
}

@article{Braun1999,
  title = {Numerical approach to solve the time-dependent Dirac equation},
  author = {Braun, J. W. and Su, Q. and Grobe, R.},
  journal = {Phys. Rev. A},
  volume = {59},
  issue = {1},
  pages = {604--612},
  numpages = {0},
  year = {1999},
  month = {Jan},
  publisher = {American Physical Society},
  doi = {10.1103/PhysRevA.59.604},
  url = {https://link.aps.org/doi/10.1103/PhysRevA.59.604}
}

@article{FillionGourdeau2012,
title = {Numerical solution of the time-dependent Dirac equation in coordinate space without fermion-doubling},
journal = {Computer Physics Communications},
volume = {183},
number = {7},
pages = {1403-1415},
year = {2012},
issn = {0010-4655},
doi = {https://doi.org/10.1016/j.cpc.2012.02.012},
url = {https://www.sciencedirect.com/science/article/pii/S0010465512000653},
author = {François Fillion-Gourdeau and Emmanuel Lorin and André D. Bandrauk}
}

@article{FillionGourdeau2014,
title = {A split-step numerical method for the time-dependent Dirac equation in 3-D axisymmetric geometry},
journal = {Journal of Computational Physics},
volume = {272},
pages = {559-587},
year = {2014},
issn = {0021-9991},
doi = {https://doi.org/10.1016/j.jcp.2014.03.068},
url = {https://www.sciencedirect.com/science/article/pii/S0021999114002630},
author = {François Fillion-Gourdeau and Emmanuel Lorin and André D. Bandrauk},
}

@article{Maquet2002,
author = {Alfred Maquet and Rainer Grobe},
title = {Atoms in strong laser fields: challenges in relativistic quantum mechanics},
journal = {Journal of Modern Optics},
volume = {49},
number = {12},
pages = {2001--2018},
year = {2002},
publisher = {Taylor \& Francis},
doi = {10.1080/09500340210140740},
URL = { 
        https://doi.org/10.1080/09500340210140740
    },
eprint = { 
        https://doi.org/10.1080/09500340210140740        
}
}

@article{Certik2013,
title = {dftatom: A robust and general Schrödinger and Dirac solver for atomic structure calculations},
journal = {Computer Physics Communications},
volume = {184},
number = {7},
pages = {1777-1791},
year = {2013},
issn = {0010-4655},
doi = {https://doi.org/10.1016/j.cpc.2013.02.014},
url = {https://www.sciencedirect.com/science/article/pii/S0010465513000714},
author = {Ondřej Čertík and John E. Pask and Jiří Vackář}
}

@article{Certik2024,
title = {High-order finite element method for atomic structure calculations},
journal = {Computer Physics Communications},
volume = {297},
pages = {109051},
year = {2024},
issn = {0010-4655},
doi = {https://doi.org/10.1016/j.cpc.2023.109051},
url = {https://www.sciencedirect.com/science/article/pii/S001046552300396X},
author = {Ondřej Čertík and John E. Pask and Isuru Fernando and Rohit Goswami and N. Sukumar and Lee. A. Collins and Gianmarco Manzini and Jiří Vackář},
}

@article{Antoine2017,
title = {Computational performance of simple and efficient sequential and parallel Dirac equation solvers},
journal = {Computer Physics Communications},
volume = {220},
pages = {150-172},
year = {2017},
issn = {0010-4655},
doi = {https://doi.org/10.1016/j.cpc.2017.07.001},
url = {https://www.sciencedirect.com/science/article/pii/S0010465517302084},
author = {X. Antoine and E. Lorin},
}

@article{Beerwerth2015,
title = {Krylov subspace methods for the Dirac equation},
journal = {Computer Physics Communications},
volume = {188},
pages = {189-197},
year = {2015},
issn = {0010-4655},
doi = {https://doi.org/10.1016/j.cpc.2014.11.008},
url = {https://www.sciencedirect.com/science/article/pii/S0010465514003804},
author = {Randolf Beerwerth and Heiko Bauke},
}

@article{Bauke2011,
title = {Accelerating the Fourier split operator method via graphics processing units},
journal = {Computer Physics Communications},
volume = {182},
number = {12},
pages = {2454-2463},
year = {2011},
issn = {0010-4655},
doi = {https://doi.org/10.1016/j.cpc.2011.07.003},
url = {https://www.sciencedirect.com/science/article/pii/S0010465511002414},
author = {Heiko Bauke and Christoph H. Keitel},
}

@article{Li2025,
title = {Robust and Scalable Federated Learning Framework for Client Data Heterogeneity Based on Optimal Clustering},
journal = {Journal of Parallel and Distributed Computing},
volume = {195},
pages = {104990},
year = {2025},
issn = {0743-7315},
doi = {https://doi.org/10.1016/j.jpdc.2024.104990},
url = {https://www.sciencedirect.com/science/article/pii/S0743731524001540},
author = {Zihan Li and Shuai Yuan and Zhitao Guan},
}

@article{Yeung2025,
title = {GPU-enabled extreme-scale turbulence simulations: Fourier pseudo-spectral algorithms at the exascale using OpenMP offloading},
journal = {Computer Physics Communications},
volume = {306},
pages = {109364},
year = {2025},
issn = {0010-4655},
doi = {https://doi.org/10.1016/j.cpc.2024.109364},
url = {https://www.sciencedirect.com/science/article/pii/S001046552400287X},
author = {P.K. Yeung and Kiran Ravikumar and Stephen Nichols and Rohini Uma-Vaideswaran}
}

@article{Sathyanarayana2025,
title = {High-speed turbulent flows towards the exascale: STREAmS-2 porting and performance},
journal = {Journal of Parallel and Distributed Computing},
volume = {196},
pages = {104993},
year = {2025},
issn = {0743-7315},
doi = {https://doi.org/10.1016/j.jpdc.2024.104993},
url = {https://www.sciencedirect.com/science/article/pii/S0743731524001576},
author = {Srikanth Sathyanarayana and Matteo Bernardini and Davide Modesti and Sergio Pirozzoli and Francesco Salvadore}
}

@article{Budiardja2023,
title = {Ready for the frontier: preparing applications for the world's first exascale system},
author = {R. Budiardja and M. Berrill and M. Eisenbach and G. Jansen and W. Joubert and D.S. Nichols and D. Rogers and A. Tharrington and B. Messer},
journal = {},
volume = {},
pages = {182–201},
year = {2023},
issn = {},
doi = {https://doi.org/10.1007/978-3-031-32041-5},
url = {https://link.springer.com/book/10.1007/978-3-031-32041-5?page=1#toc}
}

@article{Clay2018,
title = {GPU acceleration of a petascale application for turbulent mixing at high Schmidt number using OpenMP 4.5},
journal = {Computer Physics Communications},
volume = {228},
pages = {100-114},
year = {2018},
issn = {0010-4655},
doi = {https://doi.org/10.1016/j.cpc.2018.02.020},
url = {https://www.sciencedirect.com/science/article/pii/S0010465518300596},
author = {M.P. Clay and D. Buaria and P.K. Yeung and T. Gotoh},
}

@article{Friedrichs2009,
author = {Friedrichs, Mark S. and Eastman, Peter and Vaidyanathan, Vishal and Houston, Mike and Legrand, Scott and Beberg, Adam L. and Ensign, Daniel L. and Bruns, Christopher M. and Pande, Vijay S.},
title = {Accelerating molecular dynamic simulation on graphics processing units},
year = {2009},
journal = {Journal of Computational Chemistry},
volume = {30},
number = {6},
pages = {864-872},
doi = {https://doi.org/10.1002/jcc.21209},
url = {https://onlinelibrary.wiley.com/doi/abs/10.1002/jcc.21209}
}

@techreport{Asanovic2006,
    Author= {Asanović, Krste and Bodik, Ras and Catanzaro, Bryan Christopher and Gebis, Joseph James and Husbands, Parry and Keutzer, Kurt and Patterson, David A. and Plishker, William Lester and Shalf, John and Williams, Samuel Webb and Yelick, Katherine A.},
    Title= {The Landscape of Parallel Computing Research: A View from Berkeley},
    Year= {2006},
    Month= {Dec},
    Url= {http://www2.eecs.berkeley.edu/Pubs/TechRpts/2006/EECS-2006-183.html},
    Number= {UCB/EECS-2006-183},
    Institution= {EECS Department, University of California, Berkeley}
}

@misc{Top500,
  author       = {},
  title        = {TOP500 List - November 2024},
  month        = {November},
  year         = {2024},
  publisher    = {},
  doi          = {},
  url          = {https://top500.org/lists/top500/list/2024/11/},
}

@misc{GitHub,
  author       = {},
  title        = {Dirac GPU-enabled Solver},
  month        = {},
  year         = {2024},
  publisher    = {},
  doi          = {},
  url          = {https://github.com/JVembe/dirac_hydrogen_code}
}

@inproceedings{Shainer2011,
author = {Shainer, Gilad and Lui, Pak and Liu, Tong},
title = {The development of Mellanox/NVIDIA GPUDirect over InfiniBand: a new model for GPU to GPU communications},
year = {2011},
isbn = {9781450308885},
publisher = {Association for Computing Machinery},
address = {New York, NY, USA},
url = {https://doi.org/10.1145/2016741.2016769},
doi = {10.1145/2016741.2016769},
booktitle = {Proceedings of the 2011 TeraGrid Conference: Extreme Digital Discovery},
articleno = {26},
numpages = {1},
location = {Salt Lake City, Utah},
series = {TG '11}
}

@misc{Lumi,
  author       = {},
  title        = {LUMI supercomputer},
  year         = {},
  publisher    = {},
  doi          = {},
  url          = {https://www.lumi-supercomputer.eu/lumi_supercomputer/},
}

@misc{LumiGuide,
  author       = {Heikonen, Jussi and
                  Markomanolis, Georgios and
                  Achim, Cristian-Vasile and
                  Krishnasamy, Ezhilmathi and
                  Azab, Abdulrahman and
                  Ojeda-May, Pedro and
                  Martone, Michele and
                  Krotkiewski, Marcin and
                  Agueny, Hicham and
                  Barrios Sazo, Maria Guadalupe and
                  Saastad, Ole Widar},
  title        = {Best Practice Guide LUMI},
  month        = jan,
  year         = 2023,
  publisher    = {Zenodo},
  doi          = {10.5281/zenodo.7510240},
  url          = {https://doi.org/10.5281/zenodo.7510240},
}

@article{Grundmann2020,
author = {Sven Grundmann  and Daniel Trabert  and Kilian Fehre  and Nico Strenger  and Andreas Pier  and Leon Kaiser  and Max Kircher  and Miriam Weller  and Sebastian Eckart  and Lothar Ph. H. Schmidt  and Florian Trinter  and Till Jahnke  and Markus S. Schöffler  and Reinhard Dörner },
title = {Zeptosecond birth time delay in molecular photoionization},
journal = {Science},
volume = {370},
number = {6514},
pages = {339-341},
year = {2020},
doi = {10.1126/science.abb9318},
URL = {https://www.science.org/doi/abs/10.1126/science.abb9318},
eprint = {https://www.science.org/doi/pdf/10.1126/science.abb9318},
}

@article{Adnani2022,
  title = {Generation of superintense isolated attosecond pulses from trapped electrons in metal surfaces},
  author = {Adnani, Younes and Taoutioui, Abdelmalek and Makhoute, Abdelkader and T\ifmmode \mbox{\H{o}}\else \H{o}\fi{}k\'esi, K\'aroly and Agueny, Hicham},
  journal = {Phys. Rev. A},
  volume = {105},
  issue = {4},
  pages = {043104},
  numpages = {11},
  year = {2022},
  month = {Apr},
  publisher = {American Physical Society},
  doi = {10.1103/PhysRevA.105.043104},
  url = {https://link.aps.org/doi/10.1103/PhysRevA.105.043104}
}

@article{Shabaev2004,
  title = {Dual Kinetic Balance Approach to Basis-Set Expansions for the Dirac Equation},
  author = {Shabaev, V. M. and Tupitsyn, I. I. and Yerokhin, V. A. and Plunien, G. and Soff, G.},
  journal = {Phys. Rev. Lett.},
  volume = {93},
  issue = {13},
  pages = {130405},
  numpages = {4},
  year = {2004},
  month = {Sep},
  publisher = {American Physical Society},
  doi = {10.1103/PhysRevLett.93.130405},
  url = {https://link.aps.org/doi/10.1103/PhysRevLett.93.130405}
}

@article{ECormier1997,
doi = {10.1088/0953-4075/30/1/010},
url = {https://dx.doi.org/10.1088/0953-4075/30/1/010},
year = {1997},
month = {jan},
publisher = {},
volume = {30},
number = {1},
pages = {77},
author = {E Cormier and P Lambropoulos},
title = {Above-threshold ionization spectrum of hydrogen using B-spline functions},
journal = {Journal of Physics B: Atomic, Molecular and Optical Physics}
}

@article{H.Bachau_2001,
    doi = {10.1088/0034-4885/64/12/205},
    url = {https://dx.doi.org/10.1088/0034-4885/64/12/205},
    year = {2001},
    month = {nov},
    publisher = {},
    volume = {64},
    number = {12},
    pages = {1815},
    author = {H Bachau and  E Cormier and  P Decleva and  J E Hansen and  F Martín},
    title = {Applications
    of B-splines  in atomic and molecular physics},
    journal = {Reports on Progress in Physics},
}

@article{Chevalier_2008,
 author = {Chevalier, C. and Pellegrini, F.},
 title = {PT-Scotch: A Tool for Efficient Parallel Graph Ordering},
 journal = {Parallel Comput.},
 issue_date = {July, 2008},
 volume = {34},
 number = {6-8},
 month = jul,
 year = {2008},
 issn = {0167-8191},
 pages = {318--331},
 numpages = {14},
 doi = {10.1016/j.parco.2007.12.001},
 acmid = {1377241},
 publisher = {Elsevier Science Publishers B. V.},
 address = {Amsterdam, The Netherlands, The Netherlands},
}

@manual{Karypis_1998, author = {Karypis, George and Kumar, Vipin}, month = {September}, title = {METIS: A Software Package for Partitioning Unstructured Graphs, Partitioning Meshes, and Computing Fill-Reducing Orderings of Sparse Matrices}, year = 1998 }

@article{PhysRevA.37.307,
  title = {Finite basis sets for the Dirac equation constructed from B splines},
  author = {Johnson, W. R. and Blundell, S. A. and Sapirstein, J.},
  journal = {Phys. Rev. A},
  volume = {37},
  issue = {2},
  pages = {307--315},
  numpages = {0},
  year = {1988},
  month = {Jan},
  publisher = {American Physical Society},
  doi = {10.1103/PhysRevA.37.307},
  url = {https://link.aps.org/doi/10.1103/PhysRevA.37.307}
}

@article{Ivanov2015,
  title = {Relativistic calculation of the electron-momentum shift in tunneling ionization},
  author = {Ivanov, I. A.},
  journal = {Phys. Rev. A},
  volume = {91},
  issue = {4},
  pages = {043410},
  numpages = {5},
  year = {2015},
  month = {Apr},
  publisher = {American Physical Society},
  doi = {10.1103/PhysRevA.91.043410},
  url = {https://link.aps.org/doi/10.1103/PhysRevA.91.043410}
}

@article{Johanne2024.2,
doi = {10.1088/2399-6528/ad6e52},
url = {https://dx.doi.org/10.1088/2399-6528/ad6e52},
year = {2024},
month = {aug},
publisher = {IOP Publishing},
volume = {8},
number = {8},
pages = {085006},
author = {Vembe, J E and Førre, M},
title = {Anisotropic effects in the nondipole relativistic photoionization of hydrogen},
journal = {Journal of Physics Communications}
}

@article{Reiss1990,
author = {Reiss, Howard},
year = {1990},
month = {04},
pages = {574-586},
title = {Relativistic strong-field photoionization},
volume = {7},
journal = {Journal of the Optical Society of America B},
doi = {10.1364/JOSAB.7.000574}
}

@inproceedings{Gropp2000,
author = {Gropp, William and Kaushik, Dinesh and Smith, Barry and Keyes, David},
year = {2000},
month = {12},
pages = {395-404},
title = {Analyzing the Parallel Scalability of an Implicit Unstructured Mesh CFD Code},
volume = {1970},
isbn = {978-3-540-41429-2},
journal = {Lecture Notes in Computer Science},
doi = {10.1007/3-540-44467-X_36}
}

@article{Bienz2019,
title = {Node aware sparse matrix–vector multiplication},
journal = {Journal of Parallel and Distributed Computing},
volume = {130},
pages = {166-178},
year = {2019},
issn = {0743-7315},
doi = {https://doi.org/10.1016/j.jpdc.2019.03.016},
url = {https://www.sciencedirect.com/science/article/pii/S0743731519302321},
author = {Amanda Bienz and William D. Gropp and Luke N. Olson}
}

@misc{DeBoorCarl1978Apgt,
series = {Applied mathematical sciences},
volume = {27},
publisher = {Springer Verlag},
isbn = {0387903569},
year = {1978},
title = {A practical guide to splines},
language = {und},
address = {New York},
author = {De Boor, Carl},
}

@article{Antoine2017_absorb,
author = {Antoine, Xavier and Lorin, Emmanuel and Tang, Qinglin},
year = {2017},
month = {10},
pages = {1861-1879},
title = {A Friendly Review of Absorbing Boundary Conditions and Perfectly Matched Layers for Classical and Relativistic Quantum Waves Equations},
volume = {115},
journal = {Molecular Physics},
doi = {10.1080/00268976.2017.1290834}
}

@article{ANTOINE2014268,
title = {Absorbing boundary conditions for relativistic quantum mechanics equations},
journal = {Journal of Computational Physics},
volume = {277},
pages = {268-304},
year = {2014},
issn = {0021-9991},
doi = {https://doi.org/10.1016/j.jcp.2014.07.037},
url = {https://www.sciencedirect.com/science/article/pii/S0021999114005300},
author = {X. Antoine and E. Lorin and J. Sater and F. Fillion-Gourdeau and A.D. Bandrauk}
}







\end{document}